\documentclass[12pt,a4paper]{article}
\usepackage[utf8]{inputenc}
\usepackage[english]{babel}
\usepackage{amsmath}
\usepackage{amsfonts}
\usepackage{amssymb}
\usepackage{graphicx}
\usepackage[font=footnotesize,labelfont=bf]{caption}

\usepackage[colorlinks=true,citecolor=blue]{hyperref}

\usepackage{subcaption}

\usepackage{natbib}
\begin{document}

\title{Particle-in-cell simulations of the relaxation of electron beams in inhomogeneous  solar wind plasmas}

\author{Jonathan O. Thurgood
   and David Tsiklauri}
\date{} 

\maketitle

\abstract{Previous theoretical considerations of electron beam relaxation in inhomogeneous plasmas have indicated that the effects of the irregular solar wind may account for the poor agreement of homogeneous modelling with the observations.
Quasi-linear theory and Hamiltonian models based on Zakharov's equations have indicated that 
when a level of density fluctuations is above a given threshold, density irregularities act to de-resonate the beam-plasma interaction, restricting Langmuir wave growth on the expense of beam energy.
This work presents the first fully kinetic particle-in-cell (PIC) simulations of
beam relaxation under the influence of density irregularities. 
We aim to independently determine the influence of background 
inhomogeneity on the beam-plasma system, and to test theoretical predictions and alternative models using a fully kinetic treatment. 
We carry out 1D PIC simulations of a bump-on-tail unstable electron beam in the presence of increasing levels of background inhomogeneity using the fully electromagnetic, relativistic EPOCH PIC code.
We find that in the case of homogeneous background plasma density, Langmuir wave packets are generated at the resonant condition and then quasi-linear relaxation leads to a dynamic increase of wavenumbers generated.
No electron acceleration is seen - unlike in the inhomogeneous experiments, all of which produce high-energy electrons.
For the inhomogeneous experiments we also observe the generation of  
backwards propagating Langmuir waves, which is shown directly to be due to the refraction of the packets off the density gradients.
In the case of  higher-amplitude density fluctuations, similar features to the weaker cases are found, but also packets can also deviate from the expected dispersion curve in $(k,\omega$)-space due to nonlinearity. 
Our fully kinetic PIC simulations broadly confirm the findings of quasi-linear theory
and the Hamiltonian model based on Zakharov's equations. Strong density fluctuations modify properties of excited Langmuir waves altering their dispersion properties.}

\subsubsection*{}
\begin{footnotesize}

Contact:
\\jonathan.thurgood@northumbria.ac.uk
\\Department of Mathematics, Physics and Electrical Engineering, Northumbria University, Newcastle upon Tyne, NE1 8ST, UK
\\
\\d.tsiklauri@qmul.ac.uk
\\School of Physics and Astronomy, Queen Mary University of London, Mile End Road, London, E1 4NS, UK
\\
\\Authors Own (AO) copy of manuscript accepted for publication in JPP  https://doi.org/10.1017/S0022377816000970
\\The published version uses PDF-embedded movies to display time-dependent data.
\\If you have difficulties playing these movies, copies can also be found at \\https://jonathanthurgood.wordpress.com/myresearch/thurgood-and-tsiklauri-2016-movies/
\end{footnotesize}

\section{Introduction}
The processes involved in the interaction of propagating 
electron beams with background plasmas has been of long-standing 
astrophysical interest due to their close association with some 
of the most energetic solar system radio emissions, including 
Type III solar radio busts. The generally 
accepted explanation for Type III emission, the 
so-called \textit{plasma emission mechanism}, is a long-standing, 
multiple-stage model which has been considered extensively by 
numerous authors who have refined the initial 
ideas of \citet{1958SvA.....2..653G}
(although it should be noted other possibilities exist, e.g.,  such as the linear mode conversion  suggested by \citealt{1975PhRvA..11..679F,2007PhRvL..99a5003K}). In the first stage of the plasma emission model, 
electron beams which are injected in the low-corona are 
susceptible to the bump-in-tail instability as they 
propagate through the background plasma, and thus the 
beams can generate Langmuir waves. In later stages, 
these beam-generated Langmuir waves are susceptible 
to nonlinear decays and three-wave interaction processes 
which go on to produce the electromagnetic emission. 
Whilst the theoretical specifics of the later stages vary 
(see the recent review of \citealt{2014RAA....14..773S} and 
references therein), for the purposes of this paper it 
is sufficient to understand that the initial stage - the 
production of a population of Langmuir waves due to a resonant 
interaction with the beam electrons - is a common 
feature of all such models.

The most basic treatment of this stage (the beam-plasma
instability) is the application of quasi-linear 
theory (QLT) under the assumption of a homogeneous 
plasma, which describes the exchange of energy from 
diffuse, fast beam electrons into Langmuir waves at 
the resonant phase velocity (inverse Landau damping).
Once the particles reach a plateau-type distribution, 
further wave growth is prohibited, due to the absence of a
positive slope in the velocity distribution, 
and thus the instability is saturated.
Under homogeneous QLT, the calculated saturation time 
implies a flight distance of beams originating in the 
corona of only hundreds of kilometers before 
reaching saturation (e.g., \citealt{1964NASSP..50..357S}).
This is at odds with in situ solar wind measurements 
taken as far out as 1 AU, which have documented both 
the presence of bump-in-tail type electron distributions 
and growth of Langmuir waves in excess of background 
levels (see e.g. \citealt{1981ApJ...251..364L,1981JGR....86.4493A}). 
Furthermore, simple homogeneous QLT models are unable to account 
for the observed spatial clumping of Langmuir waveforms  
in the solar wind (e.g. \citealt{1978JGR....83.4147G}); the explanation of which must appeal to other processes, 
such as the kinetic localisation process described by \citet{1995JGR...10017481M,1996JGR...10115605M} 
by which the beam tends to spatially localise in a homogeneous medium as a result of nonlinearities in wave-particle resonances. Alternatively, localisations can also be produced directly as a result of background inhomogeneity, which we will explore in this paper.

Accounting for the presence of the density inhomogeneity 
in the background plasma has been shown in numerous studies to be 
a plausible explanation for the above discrepancies. 
\citet{1969JETP...30..131R}, \citet{1970JETPL..11..421B} and \citet{doi:10.1143/JPSJ.41.1757} 
considered the quasi-linear relaxation of an electron beam in the 
presence of density inhomogeneities.
They found that due to presence of the 
density inhomogeneities, a de-resonation of the 
beam-particles and background wave-field may occur, which impacts 
upon the efficacy of the beam-relaxation.
This broadening of resonances was found to be able to 
access both lower and higher velocities, and thus the 
formation of a high-energy tail in the beam velocity 
distribution function via a re-absorption of Langmuir 
wave energy was also predicted. The governing equations 
of \citet{1969JETP...30..131R} where also revisited by 
\citet{2013AnGeo..31.1379V}, who considered numerical solutions 
in order to relax assumptions about beam temperature used in 
the earlier analytical solutions.

A number of recent studies have attempted to simulate beam-relaxation 
under realistic solar wind conditions in which the 
conclusions of  \citet{1969JETP...30..131R} and \citet{1970JETPL..11..421B} 
are thought to hold, in order to determine if such a system can  
account for the aforementioned observational discrepancies.
\citet{2013ApJ...778..111K,2014JGRA..119.9369K,2015ApJ...809..176K} 
have used numerical models based on a Hamiltonian approach 
whereby the dynamics of the background plasma particles, 
Langmuir and ion-acoustic wave-fields are described by the 
Zakharov equations, which are coupled to a population of 
resonant beam-particles that are evolved according to a 
particle approach.
 
\citet{2013ApJ...778..111K} found that
the plasma inhomogeneities crucially influence the 
characteristics of the Langmuir turbulence and the beam-plasma interaction.
It was shown that the Langmuir
wave growth becomes localized and clumpy,
similar to recent observations by STEREO and other satellites. 
In their model beam particles quickly de-resonate from the Langmuir waves
and mutual energy exchange between the two is hampered much faster
than in the homogeneous plasma case.
\citet{2013ApJ...778..111K} also found that 
a tail of accelerated electrons was formed and the velocities can exceed 
significantly the beam drift velocity,  reaching two times the initial beam 
drift speed. 
\citet{2014JGRA..119.9369K} and \citet{2015ApJ...809..176K} extended 
the work, constructing observable waveforms and considering the 
interplay with nonlinear wave-wave interaction processes respectively. 
Similarly, \citet{2015ApJ...807...38V,2015JGRA..12010139V} have 
considered the specific influence of density inhomogeneities 
instead using a probabilistic model. 
They found
that for the very rapid beams with beam velocity exceeding
electron thermal speed 15 times, 
the relaxation process consists of two well-separated steps. 
In the first step the major relaxation process occurs with 
the wave growth rate everywhere becoming almost close to zero or negative. 
In the second stage the system remains in 
the state close to state of marginal stability 
long enough to explain how the beam may be preserved 
travelling distances over 1 AU while still being 
able to generate the Langmuir waves that are 
usually detected in-situ by satellites. 

In light of these recent results there is a growing 
consensus that inhomogeneity will significantly modify 
the beam-plasma system and so cannot be ignored in the 
context of electron beams propagating in the solar wind. 
However, the changes to the relaxation of electron beams 
due to plasma inhomogeneity are yet to be considered 
using a fully-kinetic approach. 
There are a two main motivations for considering the problem within a fully kinetic particle-in-cell (PIC) approach. Firstly, the PIC description of plasmas is in principle the most generalised approach and can describe a richer physical picture (theoretically containing all of the physics, in the limit of realistic numbers of particles). Secondly, the use of beam-only kinetics imposes an upper limit on excitable wave vectors as the beam relaxes towards smaller velocities during the velocity-space plateau formation. Thus, we note that the fully kinetic PIC simulation has the advantage of accessing the full $k$-range.
In this paper we therefore 
present the results of the first such study whereby an
electron beam's interaction with an inhomogeneous background 
is simulated using the fully kinetic particle-in-cell method.

The paper is structured as follows: in Section \ref{setup} 
we describe the parameters used in our simulations.
In Section \ref{homo_results} we present the results for the 
homogeneous experiment, followed by the results of the 
different inhomogeneous results in Section \ref{inhomo_results}. 
The implications of the results are discussed in 
Section \ref{discussion}, and conclusions are drawn 
in Section \ref{conclusion}.

\section{Numerical Setup}\label{setup}

\begin{figure}
  \centerline{\includegraphics[width=13cm]{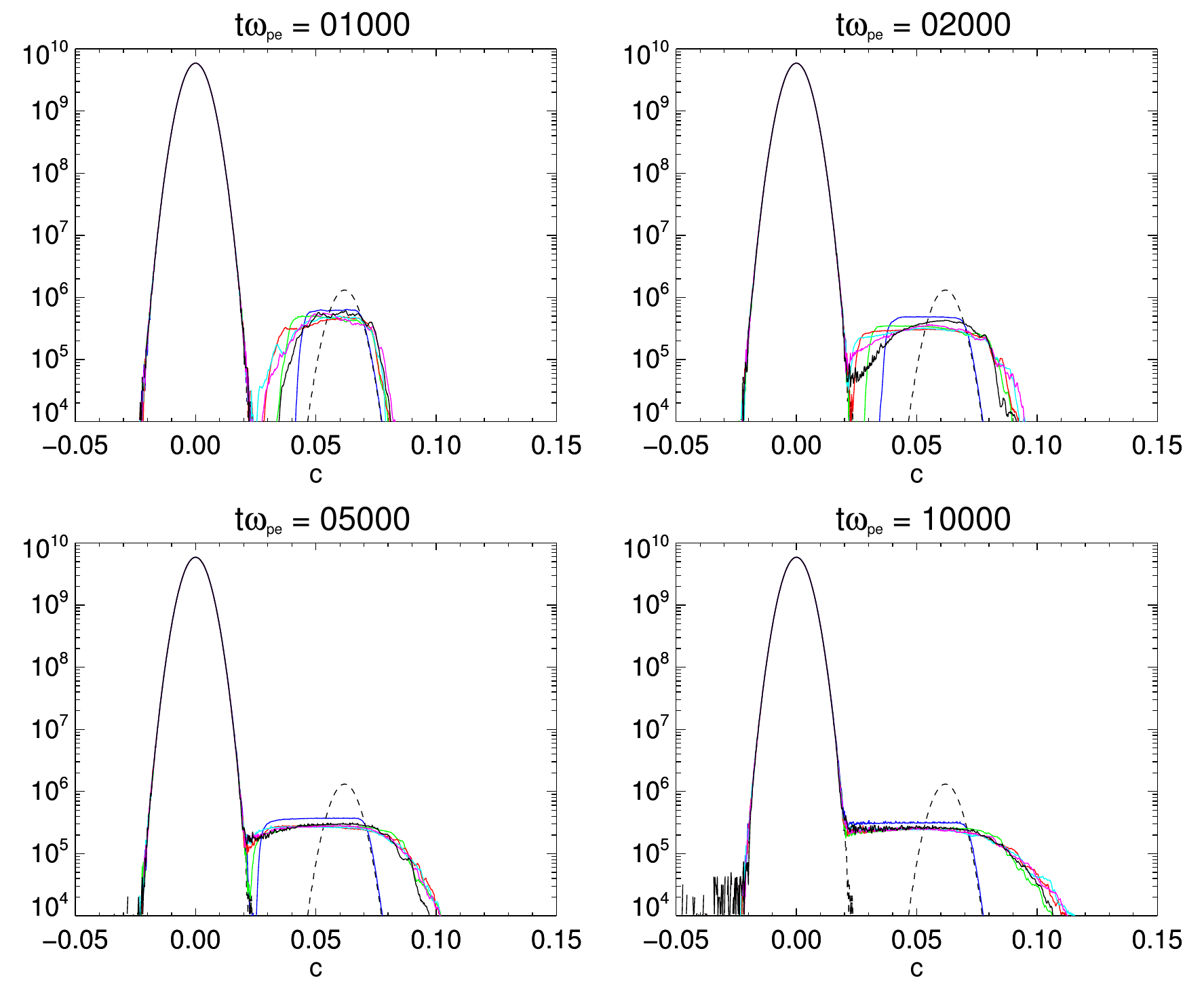}}
       \caption{
     Influence of the level of inhomogeneity on the evolution of the 
     electron velocity distribution function with time, 
     where  ${\Delta}n = 0$ (blue), $0.01$ (green), $0.02$ (red), 
     $0.03$ (cyan), $0.04$ (magenta) and $0.05$ (black). 
     The initial profile of the electrons (common to all cases) 
     is shown by the black dashed line. }
     \label{fig_vdf}
  
\end{figure}

In this paper we present the results of six numerical 
experiments in which a bump-in-tail unstable electron 
beam is initialized in the presence of a Maxwellian 
background which is increasingly irregular in space. 
The simulations are carried out in a one-dimensional (1D) 
geometry using EPOCH, an open-source multidimensional 
particle-in-cell code. Full details of the underlying solver, 
along with benchmarking on test 
problems, are available in \citet{Arber:2015hc}.

All parameters except for those relating to the spatial profile 
of the background particles are common to all six experiments. 
The homogeneous background plasma parameters, upon which the density 
irregularities are then imposed, is chosen as follows: 
the homogeneous background number density 
$n_{0} = n_{e} = n_{i} = 5\times10^{6}\, \mathrm{m}^{-3}$, 
and background electron temperature $T_{e} = 10\,\mathrm{eV}$ 
The beam particles are 
initialised with a shifted Maxwellian profile, with density 
$n_{b}/n_{0} = 2.5\times10^{-4}$, speed $v_{b} = 14 V_{th,e}$ and 
velocity spread of ${\Delta}v_{b}=0.08v_{b}$. The initial electron 
velocity distribution function may be seen in the 
dashed-curves of Figure \ref{fig_vdf}. 

The background density profile $n_0$ is modified from 
simulation-to-simulation with the addition of a 
spatially-dependent perturbation $\delta n\left(x\right)$  such 
that $n_{0}^{\prime}\left(x\right) = n_{0} + {\delta}n(x)$. The 
perturbation's profile is comprised of ten harmonics:

\begin{equation}\label{profile_eqn}
\delta n(x) = \frac{{\Delta}n}{N} \, \sum\limits_{i=1}^{10} A_{i}
\mathrm{\sin}\left(\frac{2{\pi}x + 2{\pi}{\phi}_i}{\lambda_i} \right)\,.
\end{equation}

Each harmonic's parameters were chosen randomly from a 
uniform distribution ranging from $0-1$ (amplitude $A_{i}$), 
$300\lambda_{D} - 2000\lambda_{D}$  (wavelength $\lambda_{i}$) 
and  $0-1$ (phase-shift $\phi_{i}$). The resulting signal was 
then normalised in amplitude by a factor $N$ that is 
calculated such that the average amplitude of the profile in 
question ${\delta}n(x)$ can be fixed by a choice of ${\Delta}n$ such that 
${\Delta}n=\left\langle\left({\delta}{n}/n_{0}\right)^{2}
\right\rangle^{\frac{1}{2}}$. Thus, the experiments considered 
vary as a choice of average profile percentage density fluctuation. 

Following \citet{1969JETP...30..131R}, the modified 
Langmuir wave dispersion relationship in the presence 
of such a density profile is to the first order
\begin{equation}\label{first_order_eqn}
\omega\left(k,x\right) \approx
\omega_{pe}\left(
1 + \frac{1}{2}\frac{{\delta}n(x)}{n_{0}}
\right)
+ \frac{3}{2}\frac{k^{2}V_{th,e}^{2}}{\omega_{pe}} \,.
\end{equation}
Where $\omega_{pe}$ is the electron plasma frequency and $V_{th,e}$ the electron thermal velocity. Thus, for density fluctuation to be large enough
to cause beam-Langmuir wave de-resonance
we require the density ratio term on right-hand-side of the
equation (\ref{first_order_eqn}) to be larger than the third term. This yields
\begin{equation}
\Delta n\geq 3\frac{k^{2}V_{th,e}^{2}}{\omega_{pe}^{2}}.
\end{equation}
Noting that the expected wavenumber associated with 
the two stream instability is approximately $k\approx\omega_{pe}/v_{b}$ 
then we deduce
\begin{equation}\label{density_threshold}
\Delta n\geq 3\left(\frac{V_{th,e}}{v_b}\right)^2 \,.
\end{equation}
We therefore expect choices of average density profile 
satisfying equation (\ref{density_threshold}) to act in the 
inhomogeneous regime described by \citet{1969JETP...30..131R}. 
In this paper we present results for choices $\Delta{n} = 0$ 
(a homogeneous plasma), and ${\Delta}n = 0.01,0.02,...,0.05$ 
(inhomogeneous plasmas which are in excess of the threshold 
equation (\ref{density_threshold}) as calculated for our 
parameter choices).
 The resulting density profiles for 
our specific choices of ${\Delta}n$ are illustrated 
later in Figures \ref{packet_0pc} and 
Figures \ref{packet_1pc}--\ref{packet_5pc}.

At this stage it is important to note two features 
of this  profile. Firstly, we immobilise the ions 
and thus keep this profile fixed throughout the  
simulation.
Thus, the ion-to-electron mass ratio is infinite and results are independent of the ion temperature.
This is necessary in that it avoids 
the problem of the fluctuations subsiding due to kinetic 
damping in the absence of an external pump driving 
ion-acoustic turbulence, which in test runs was 
found to occur much more rapidly than the relaxation 
time of the instability (and so, otherwise mobile 
ions would prohibit the experiment).  Secondly, it 
follows that the profile's spatial scale is in 
fact fixed (the amplitude of each \lq{potential well}\rq{} 
is a variable, however the length is fixed). This allows 
for direct comparisons regarding the specific influence 
of the amplitude as opposed to that of changing spatial 
scales which, according to inhomogeneous QLT, are also a 
factor in influencing the relaxation 
\citep[see, e.g.,][]{1969JETP...30..131R}.
The peak-to-peak spatial scales of the resulting density cavities (arising from our choice of constituent harmonics in equation \ref{profile_eqn})  are around $1000 \lambda_D$, two orders of magnitude longer than the expected wavelengths excited by the beam-plasma interaction, which in this case is approximately $14 \lambda_D$. Thus, this satisfies two (intuitive) scale requirements of the inhomogeneity; namely that scale should be larger than the Langmuir wavelengths so that the waves fit within the cavities, but be sufficiently short compared to the wave propagation scale (the product of their group speed and growth rate) such that the waves will experience refractive effects  before the saturation of the instability (viz. the gradient is not too gentle). 
Such requirements are formally discussed by  \citet{1969JETP...30..131R}.

The chosen beam and (homogeneous) background parameters, 
except for the beam-to-background density ratio, are the same 
those used in the aforementioned Hamiltonian model of  
\cite{2013ApJ...778..111K}. This choice is motivated by both 
suitability for solar wind plasmas and also to maximise 
comparability between the results. However, 
because the quasilinear instability time $\tau_{ql}$ is known to scale as 
$\tau_{ql}\omega_{pe} =(n_b/n_0)^{-1}(v_{b}/{\Delta}v_{b})^2 $ (see \citealt{9780511600036}),  
running a fully kinetic PIC code for a diffuse astrophysical beam 
(where $n_b/n_0 \sim 10^{-6}$, i.e., running for hundreds-of-thousands 
of plasma periods) with sufficient resolution, and particle 
representation and time-step (fractions of a plasma period) 
is unfeasible. As such, we have taken $n_b/n_0$ as an order 
of magnitude larger in order to be able to complete the 
simulations with the available computational resources. As discussed 
by \cite{2015A&A...584A..83T}, due to the beam-plasma 
system's sensitive parameter space, this may compromise 
complete comparability.

The simulations are ran for $10^{4}$ plasma periods 
($\omega_{pe}^{-1}$) in a periodic domain of size 
$L_{x} = 5000 \lambda_D$ with  5000 cells, thus grid is 
resolved to the Debye length ${\Delta}x = \lambda_D$ and 
comfortably spans many individual cavities of the density 
profile. 
With a beam velocity of $v_{b}=14V_{th,e}$, streaming beam electrons will cross the grid around $28$ times during the simulation, crossing through many individual density cavities and being recycled through the boundaries. The expected group velocity of waves excited by the two stream instability is much slower - of order $\sim 0.2V_{th,e}$ , and so (unmodified) propagating wave packets will travel distances of around $2000 \lambda_D$, in excess of the cavity lengths, and so should experience the effects of the density gradient.
The number of computational particles per cell 
per species (PPCPS) used is $5000$ for background electrons 
and ions, and $1315$ PPCPS for the beam electrons. The cell-to-cell particle loading noise associated with the choice of $5000$ PPCPS  for the background has a RMS fluctuation of around $0.3\%$, an order of magnitude less than inhomogeneous density threshold (\ref{density_threshold}). Thus, the inhomogeneous density profiles considered in this paper are clearly resolved against the noise. This particle resolution was the maximum given available resources, and as such we were unable to confidently separate intermediate profiles where $0<{\Delta}n<3\left(V_{th,e}v_{b}^{-1}\right)^{2}$  from the noise levels - thus we do not consider the intermediate case of \lq{weak inhomogeneity}\rq{} in this paper.
Each run took approximately 72 hours of run-time on 
128 cores (8x2x8 core Intel Xeon E5-2650 v2 processors).

\section{Results}
\subsection{Homogeneous regime}\label{homo_results}

\begin{figure}
  \centerline{\includegraphics[width=9cm]{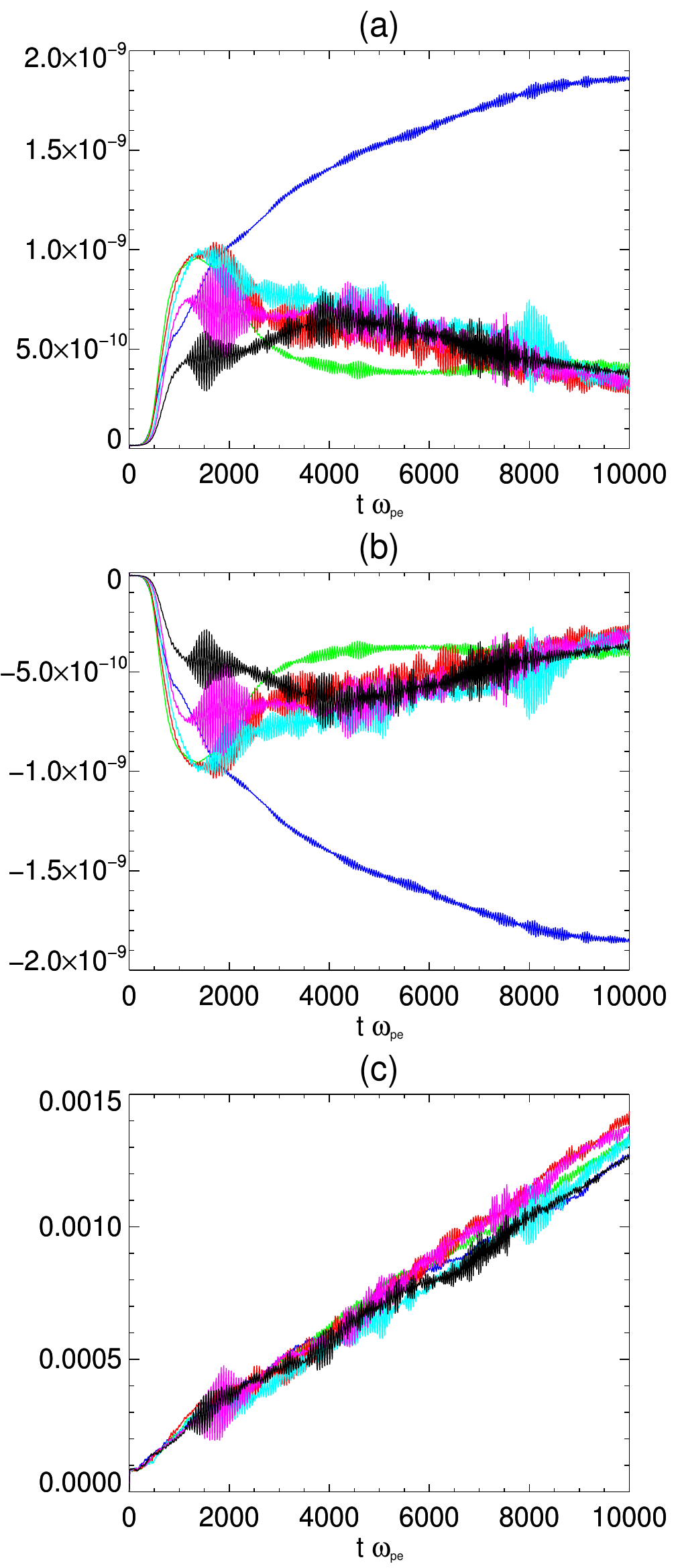}}
  \caption{Evolution of (a) total field energy ($J$), 
  (b) total particle energy ($J$), and (c) total simulation 
  energy ($\%$ change) over time for the six runs of 
  increasing inhomogeneity amplitude - ${\Delta}n = 0$ (blue), 
  $0.01$ (green), $0.02$ (red), $0.03$ (cyan), $0.04$ 
  (magenta) and $0.05$ (black). }
  \label{fig_energy}
\end{figure}

\begin{figure}
  \centerline{\includegraphics[width=9cm]{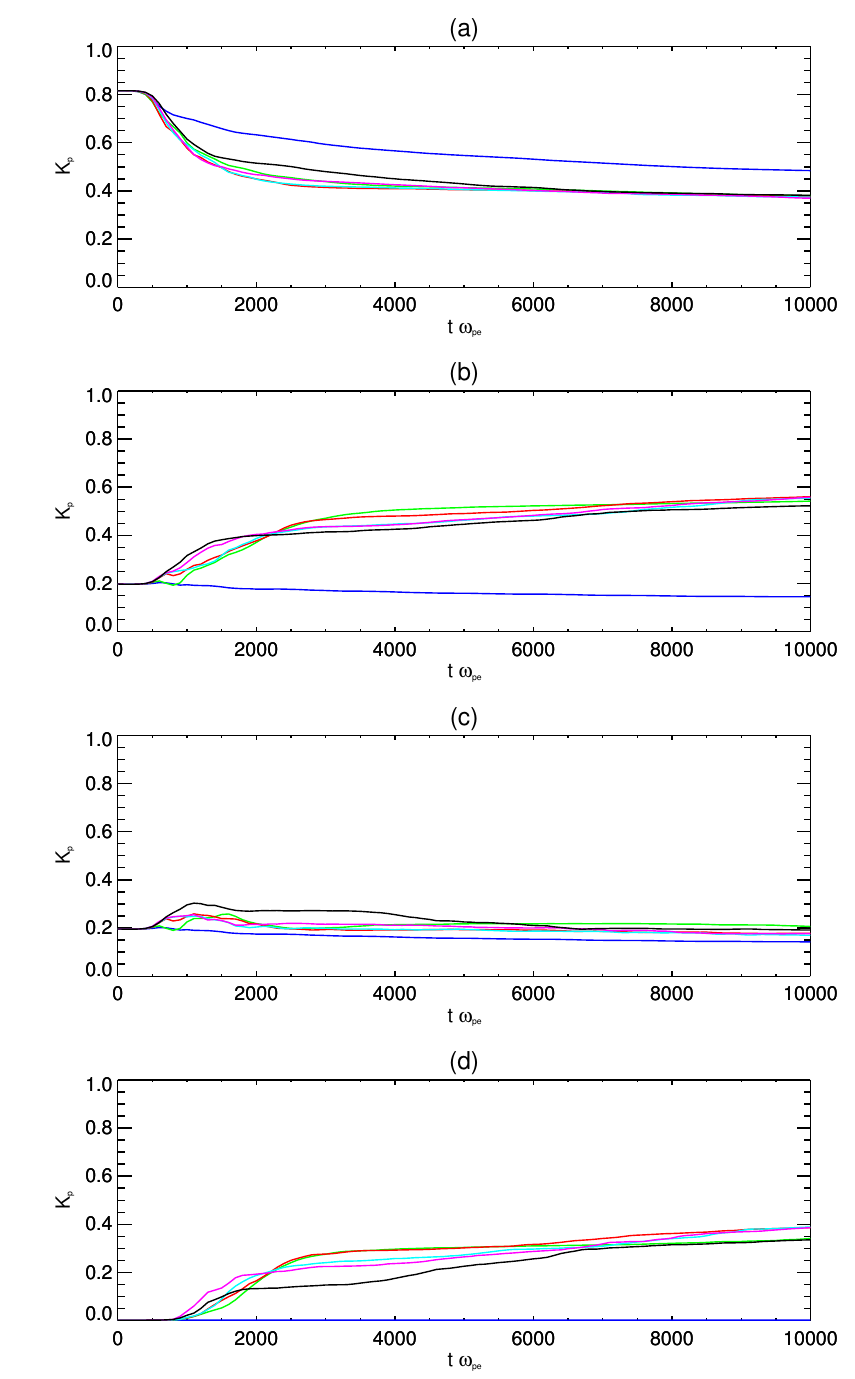}}
  \caption{
Evolution of the kinetic energy budget of the beam 
particles in four different velocity ranges for 
different values of ${\Delta}n$, showing net 
acceleration and deceleration processes.  
The four velocity bands correspond to 
$K_{p}\left(v<v_{b}+{\Delta}v_{b}\right)$ (i),
$K_{p}\left(v>v_{b}+{\Delta}v_{b}\right)$ (ii),
$K_{p}\left(v_{b}+{\Delta}v_{b}<v<v_{b}+3{\Delta}v_{b}\right)$ (iii)
and $K_{p}\left(v>v_{b}+3{\Delta}v_{b}\right)$ (iv).  
$K_{p}$ is normalised by the initial kinetic energy of the beam.
The level of inhomogeneity is indicated by line color -  
${\Delta}n = 0$ (blue), $0.01$ (green), $0.02$ (red), 
$0.03$ (cyan), $0.04$ (magenta) and $0.05$ (black).
  }
     \label{fig_kebands}
\end{figure}

\begin{figure}
  \centerline{\includegraphics[width=11cm]{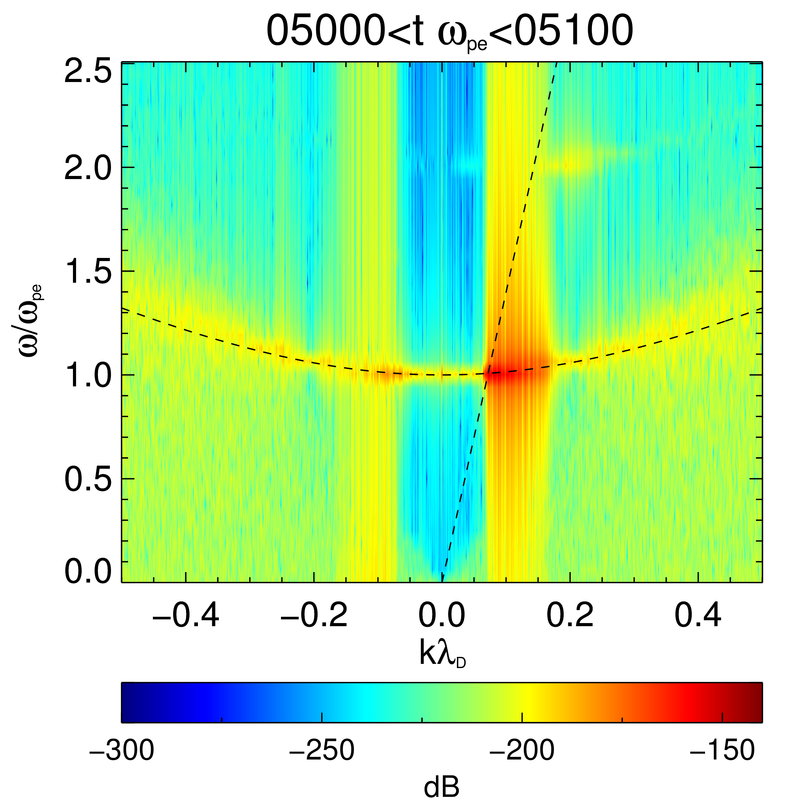}}
  \caption{
  Embedded movie showing the evolution of time-windowed Fast Fourier Transforms of electric field energy in $\left(k,\omega\right)$-space for the homogeneous simulation. 
  \textbf{https://jonathanthurgood.wordpress.com/myresearch/thurgood-and-tsiklauri-2016-movies/}  }
  \label{kw_0pc}
\end{figure}

\begin{figure}
  \centerline{\includegraphics[width=11cm]{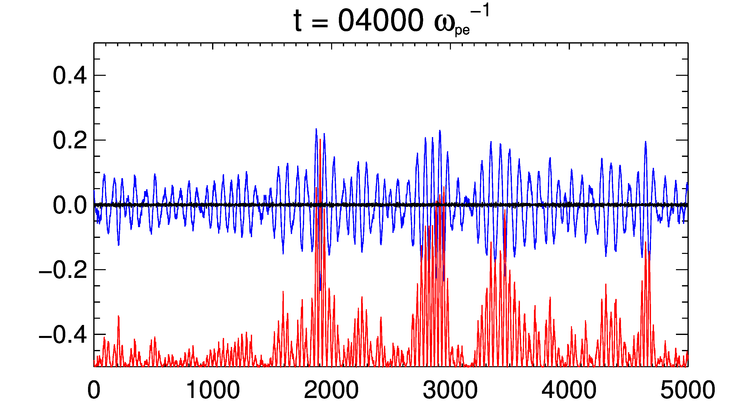}}
  \caption{
  Embedded movie showing the wave packet evolution for ${\Delta}n=0$. 
  The blue curve shows the spatial profile of the 
  electric field $E_{x}$ ($\mathrm{V}/\mathrm{m}$), 
  the red shows $10E_{X}^{2}-0.5$, and the black 
  shows the density profile ${\delta}n/n_{0}$. 
  \textbf{https://jonathanthurgood.wordpress.com/myresearch/thurgood-and-tsiklauri-2016-movies/}    
  }
  \label{packet_0pc}
\end{figure}

We begin by describing the results of the 
homogeneous (${\Delta}n=0$) simulation before comparing 
with results for the inhomogeneous regime 
(where 
${\Delta}n$ satisfies equation \ref{density_threshold}) 
in Section \ref{inhomo_results}.

Figure \ref{fig_vdf} shows the evolution of the electron  
velocity distribution function of the six runs of 
increasing inhomogeneity ${\Delta}n = 0,0.01,0.02,...,0.05$ 
at four instances in time ($t{\omega}_{pe} = 
1000, 2000, 5000, 10000$). 
Focusing on the ${\Delta}n=0$ case for now (blue curve) 
we see the saturation of the initial bump-in-tail 
instability and the merging of the beam electrons to the 
bulk distribution, resulting in the characteristic plateau formation. 
This is observed to occur by around $t\omega_{pe} = 9000$, 
which is in excess of the time scale predicted by the aforementioned quasi-linear theory 
formula 
($\tau_{ql}\omega_{pe} =(n_b/n_0)^{-1}(v_{b}/{\Delta}v_{b})^2 $). 
We however, confirmed that this is a realistic relaxation 
time for this setup in our convergence testing, 
rather than being artificially hastened as a 
side-effect of poor computational particle counts 
as reported by \cite{2014PhPl...21l2104R} and \cite{2015PhPl...22b4502L}. 
The discrepancy is instead due to the use 
of the relatively dense and energetic beams which 
are outside of the formal applicability of the 
quasi-linear theory, as discussed in Section \ref{setup}. 
Furthermore, we note that there is no formation 
of a high-energy tail which is consistent with 
expectations of homogeneous quasi-linear theory. 
\cite{2014PhPl...21l2104R} and \cite{2015PhPl...22b4502L} 
have both also noted that insufficient PPCPS leads to 
the erroneous formation of an unphysical high-energy tail. 
As such, this serves as confirmation that our PPCPS count 
is sufficient to correctly resolve the beam-plasma 
instability with no such numerical artefact. Given 
that there is no such artefact for ${\Delta}n=0$, in 
Section \ref{inhomo_results}, when we consider inhomogeneous results,  
any sign of high-energy tail formation is therefore 
physically realistic. 

The evolution of total field and particle energy is 
shown in Figure \ref{fig_energy} for the various simulations.
For ${\Delta}n=0$ we see the instability onset at around 
$t\omega_{pe}=250$
which corresponds to a growth of electric field energy 
in panel (a) and a concomitant loss of particle energy in panel (b). 
The growth continues until the saturation time near 
$t\omega_{pe} = 9000$,  which corresponds to the 
observed plateau formation time in Figure \ref{fig_vdf}. In panel (c), 
we see that the total simulation energy is well-conserved 
to within $<0.002\%$ (for all of the simulations, 
i.e. numerical heating is negligible). 
Figure \ref{fig_kebands} more specifically shows the 
kinetic-energy evolution of the population of beam particles, 
contained within different velocity ranges. 
Specifically, the four bands are (i) 
$K_{p}\left(v<v_{b}+{\Delta}v_{b}\right)$,
(ii) $K_{p}\left(v>v_{b}+{\Delta}v_{b}\right)$,
(iii) $K_{p}\left(v_{b}+{\Delta}v_{b}<v<v_{b}+3{\Delta}v_{b}\right)$ 
 and 
 (iv) $K_{p}\left(v>v_{b}+3{\Delta}v_{b}\right)$. 
 In the ${\Delta}n=0$ case (blue curve) we see that 
 the beam has a net loss of kinetic energy across the 
 different ranges, primarily from the main thermal bulk in range (i). 
There is no gain of kinetic energy in any range which 
is consistent with the absence of any high energy tail 
formation for ${\Delta}n=0$ in Figure \ref{fig_vdf}. 

We now turn our attention to the behaviour of the 
electrostatic waves generated by the beam relaxation. 
Figure \ref{kw_0pc} shows the evolution of electric field 
energy of the homogeneous simulation in $\left(k,\omega\right)$-space 
by considering 2D, Fourier transforms in $(x,t)$-space 
windowed over subsequent time periods. The time window size of 
$100\,\omega_{pe}^{-1}$ was determined by experimentation 
and provides the best balance between frequency 
resolution and effective time cadence, allowing us 
to track the evolution in both Fourier space and time, 
whilst preserving sufficient frequency resolution to 
reasonably compare spectra to expected dispersion curves. 
The decibel scale is defined with a reference (spectral) energy density level of  1, i.e., the plotted quantity is $10 \log \left(0.5\epsilon_{0}\mathcal{F}(E_{x}\left(x,t\right))^{2}\right)$.
The dispersion curves for the expected beam mode 
$\omega = v_{b}k $
and the (unmodified by the beam) Langmuir mode 
$\omega^{2} = \omega_{pe}^{2} + 3k^{2}V_{th,e}^{2}$
are overlaid.
We find that the majority of initial growth occurs on 
the intersection of the beam-mode and Langmuir mode, i.e., 
the expected wave-particle resonance. As time advances, 
this spectral energy density shifts to the right of the 
intersection of the overlaid curves \textit{(i.e., the 
resonance shifts along the Langmuir dispersion curve 
towards higher k)}.
This is due to a dynamic shift in the resonance 
point as the instability proceeds - as beam particles 
loose energy and join the main population, the effective 
beam velocity is reduced and so the resonance point 
shifts along the Langmuir wave dispersion curve to higher $k$.
There is also some development of spectral energy 
density in the negative $k$ for the homogeneous case 
(see especially blue curve in  Figure \ref{fig_kspe}). 
This is a minor effect, barely noticeable in 
Figure \ref{kw_0pc}. We also note the presence of a spectral peak at the harmonic of the beam-driven waves, i.e., at $\left(2k_{L},\omega_{L}\right)$. This could be a signature of kinetic localisation as described by \citet{1995JGR...10017481M,1996JGR...10115605M}. 

Finally, with regards to the spectra, we also note the 
presence of a small spectral peak at $k = 0$ visible at 
early times. This is most clearly visible in Figure \ref{fig_kspe},  
where the spectra of Figure \ref{kw_0pc} and similar 
have been integrated over $\omega$ for direct comparison between 
different choices of ${\Delta}n$ (see Section 
\ref{inhomo_results} for full details). Occurring for all 
values of ${\Delta}n$, this corresponds to a beam-aligned, 
standing mode of the electric field $E_{x}$ oscillating at 
the local Langmuir frequency, which is present from the 
simulation initialisation (thus, before the instability onset). 
It is caused by the non-zero initial current imposed by the 
beam at t = 0, and has been discussed in detail by 
\citet{2013AnGeo..31..633B}. Whilst it is possible to 
remove this mode by introducing a compensating drift 
velocity to the background electrons, we tolerate its 
presence as it is unclear whether such a compensation 
is physically appropriate (in particular, it may influence 
the correct return-current processes). Regardless, we have 
found that the amplitude associated with this mode, is in 
all runs much less than that of the Langmuir waves generated 
after the instability onset and as such it does not  
significantly affect the dynamics of the participating 
Langmuir wave population. 

The time evolution of the generated electrostatic waves is
presented in 
Figure \ref{packet_0pc}. In particular,  it shows the 
evolution of the electric field in $(x,t)$-space in the homogeneous case 
(blue curve), and a scaled-measure of electric field energy 
density ($10E_{X}^{2}-0.5$, red-curve). After the instability 
onset, a near-monochromatic wave, with a clear rightward-propagating 
character starts to grow.
In the time interval $t\omega_{pe}=2000-4000$, we note two effects in the homogeneous simulation - firstly, a coherent amplitude modulation 
due to the beating with the aforementioned standing 
oscillation of the electric field (see also \citealt{2013AnGeo..31..633B}), and secondly, signs of  amplitude localisation in 
$E_{x}$, or \lq{clumping}\rq{}. The localisation in this case, given the absence of both density inhomogeneity and ion-acoustic participation (due to fixed ions), is due to the nonlinear kinetic localisation as described by \cite{1995JGR...10017481M,1996JGR...10115605M}, a process which is consistent also with the observation of the harmonic signal previously noted in the Fourier spectra.

\subsection{Inhomogeneous regime and comparisons}\label{inhomo_results}

\begin{figure}
  \centerline{\includegraphics[width=11cm]{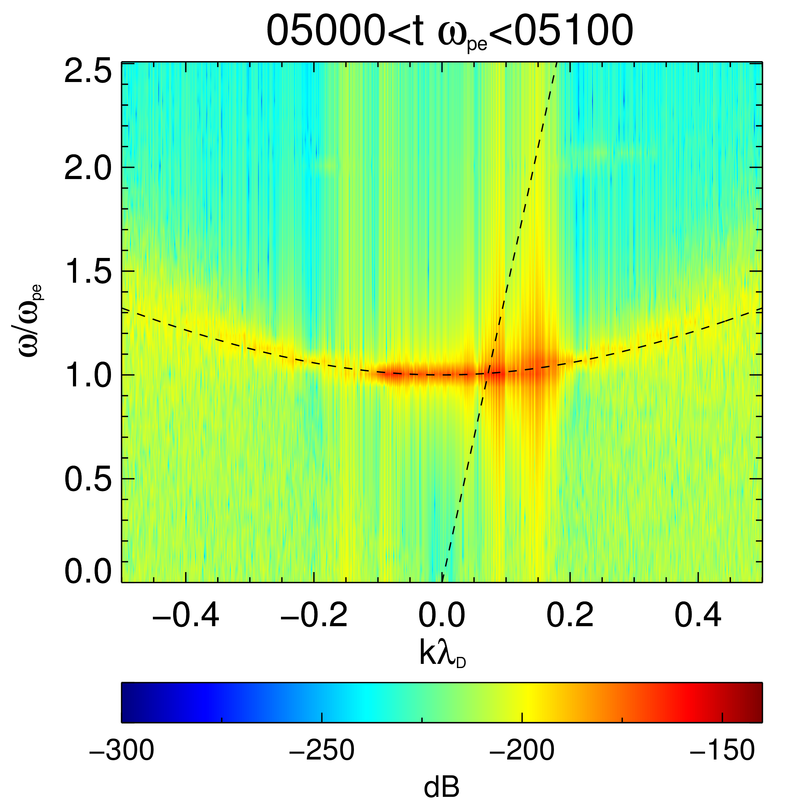}}
  \caption{
  Embedded movie showing the evolution of time-windowed Fast Fourier Transforms of electric field energy in $\left(k,\omega\right)$-space for the ${\Delta}n=0.01$ simulation. 
  \textbf{https://jonathanthurgood.wordpress.com/myresearch/thurgood-and-tsiklauri-2016-movies/}  }

  \label{kw_1pc}
\end{figure}

\begin{figure}
  \centerline{\includegraphics[width=11cm]{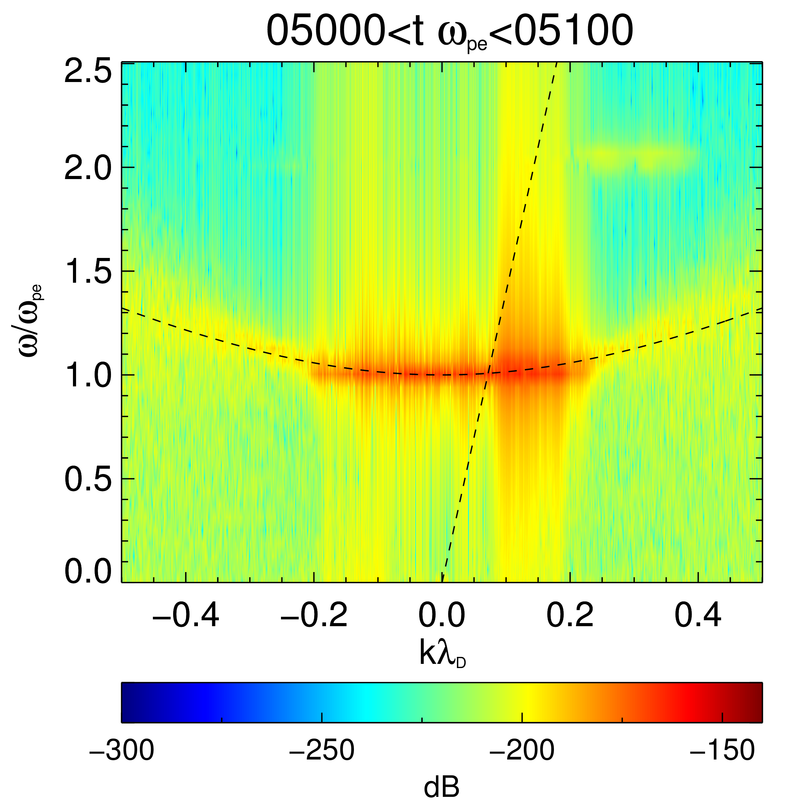}}
  \caption{
  Embedded movie showing the evolution of time-windowed Fast Fourier Transforms of electric field energy in $\left(k,\omega\right)$-space for the ${\Delta}n=0.05$ simulation. 
  \textbf{https://jonathanthurgood.wordpress.com/myresearch/thurgood-and-tsiklauri-2016-movies/}  }
  \label{kw_5pc}
\end{figure}

\begin{figure*}
  \centerline{\includegraphics[width=13cm]{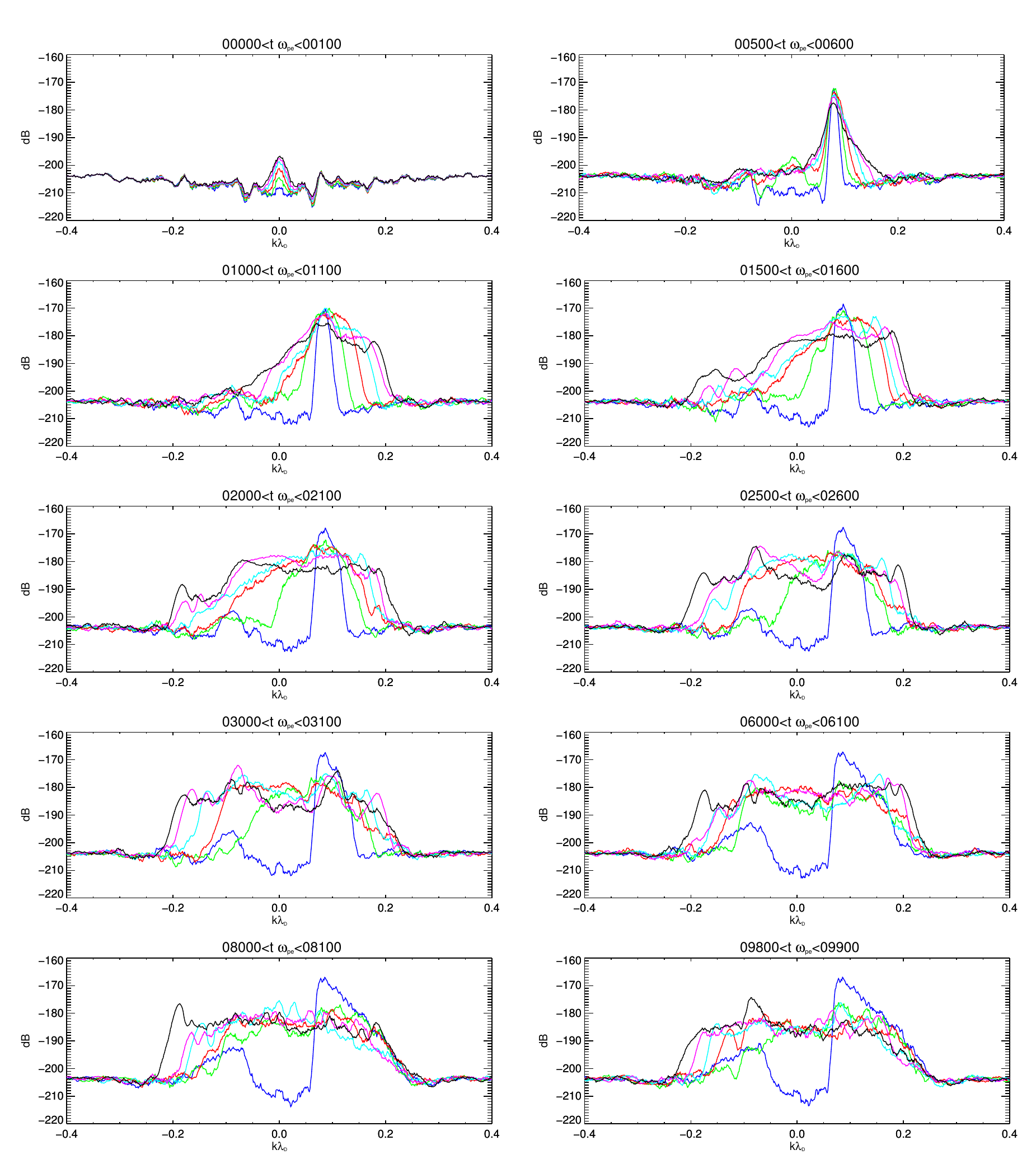}}
    \caption{
Comparison of $k$-space evolution of electric field energy for 
different levels of inhomogeneity with time, where  
${\Delta}n = 0$ (blue), $0.01$ (green), $0.02$ (red), 
$0.03$ (cyan), $0.04$ (magenta) and $0.05$ (black).
     }
     \label{fig_kspe}
\end{figure*}

\begin{figure}
  \centerline{\includegraphics[width=11cm]{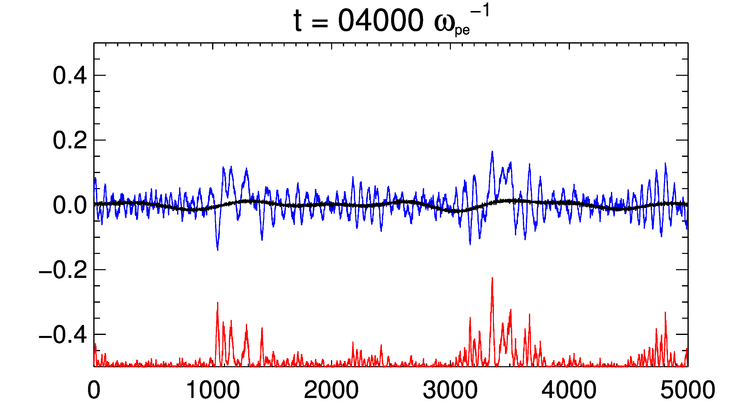}}
  \caption{
  Embedded movie showing the wave packet evolution for ${\Delta}n=0.01$. 
  The blue curve shows the spatial profile of the 
  electric field $E_{x}$ ($\mathrm{V}/\mathrm{m}$), 
  the red shows $10E_{X}^{2}-0.5$, and the black 
  shows the density profile ${\delta}n/n_{0}$. 
  \textbf{https://jonathanthurgood.wordpress.com/myresearch/thurgood-and-tsiklauri-2016-movies/}    
  }

  \label{packet_1pc}
\end{figure}

\begin{figure}
  \centerline{\includegraphics[width=11cm]{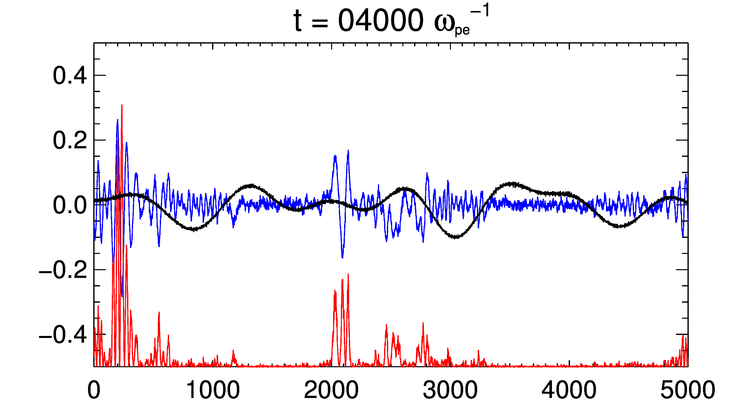}}
  \caption{
  Embedded movie showing the wave packet evolution for ${\Delta}n=0.05$. 
  The blue curve shows the spatial profile of the 
  electric field $E_{x}$ ($\mathrm{V}/\mathrm{m}$), 
  the red shows $10E_{X}^{2}-0.5$, and the black 
  shows the density profile ${\delta}n/n_{0}$. 
  \textbf{https://jonathanthurgood.wordpress.com/myresearch/thurgood-and-tsiklauri-2016-movies/}    
  }

  \label{packet_5pc}
\end{figure}

\begin{figure}
  \centerline{\includegraphics[width=13cm]{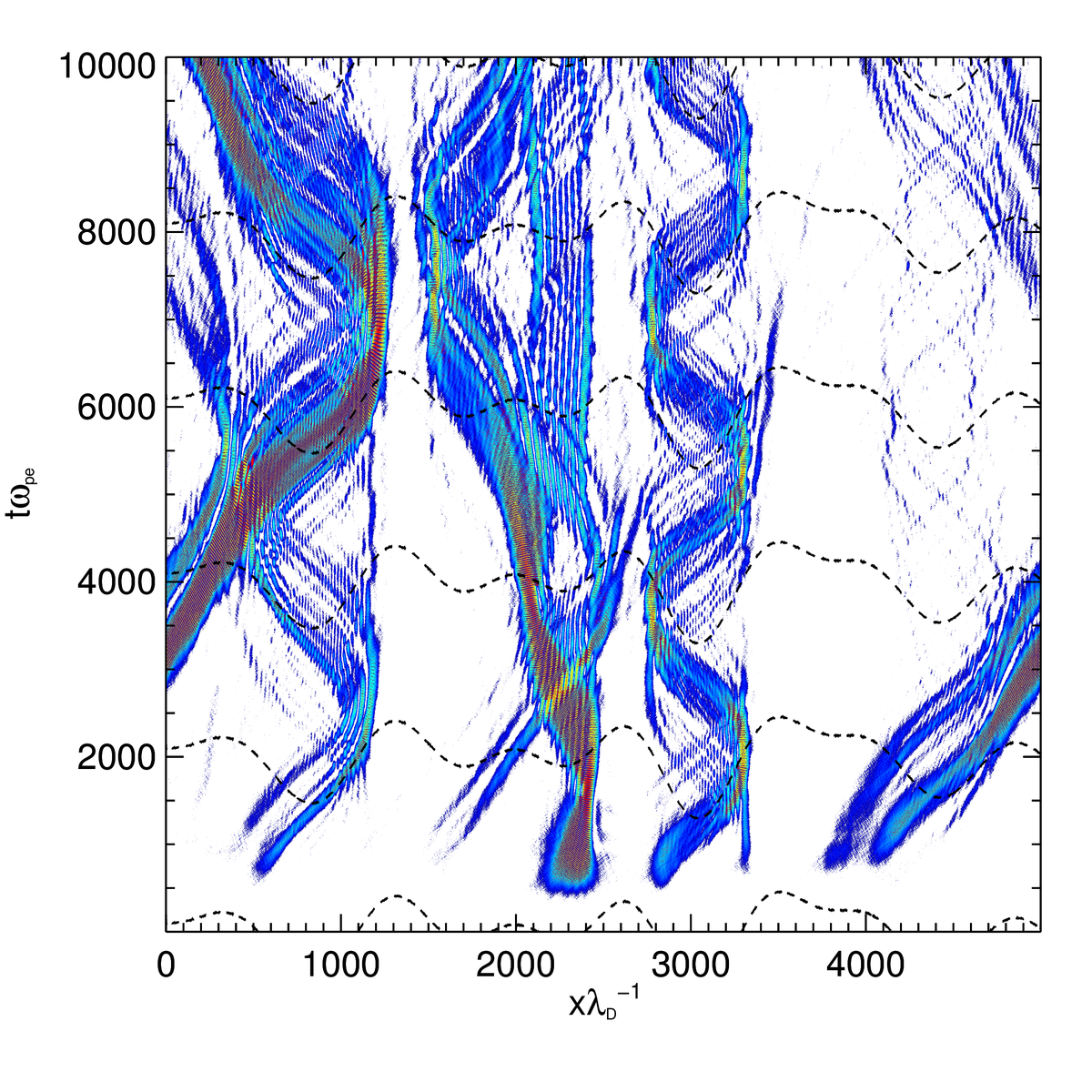}}
  \caption{
  Time-distance diagram of electric field energy for ${\Delta}n=0.05$, illustrating the temporal evolution of the spatially localised wave packets. Note the reflections of the packets between the walls of the density cavities, where the black-dashed lines illustrate the density profile along $x$. Levels of the colour table are logarithmically spaced, for best contrast between features (see Figure \ref{packet_5pc} for corresponding quantitative information).
  }
  \label{td_5pc}
\end{figure}

\begin{figure}
  \centerline{\includegraphics[width=13cm]{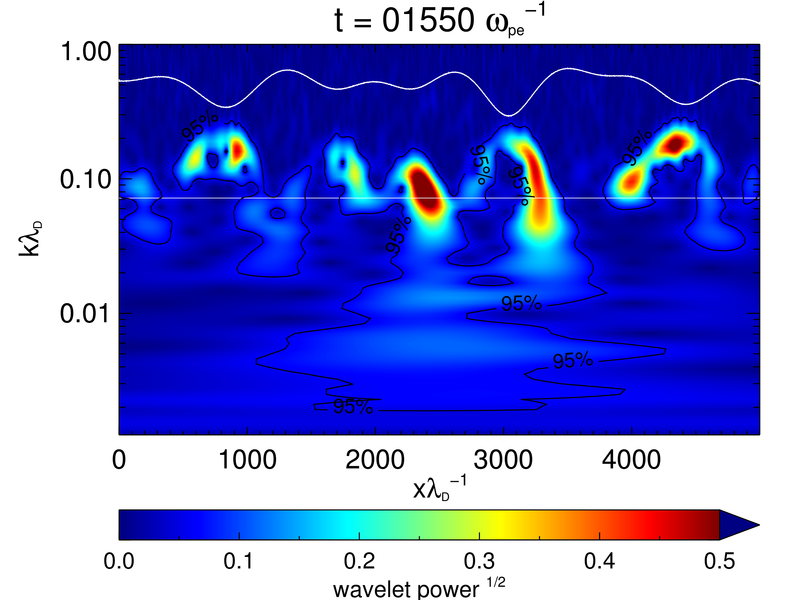}}
  \caption{
    Embedded movie of wavelet power spectrum of the electric field in $\left(x,k\right)$-space for  ${\Delta}n=0.05$. The localisation of wave packets on density gradients is further illustrated, as is the kinematic effect of the reduction in $k$ during the reflection on positive gradients (cf., Figure \ref{td_5pc}). The white curve illustrates the density profile in $x$, and the straight white line is the homogeneous beam-particle resonance $k{\lambda_{D}}=0.072$. Note the \lq{bursty}\rq{} behaviour of the low-$k$ modes. 
   \textbf{https://jonathanthurgood.wordpress.com/myresearch/thurgood-and-tsiklauri-2016-movies/}  
    }
  \label{wavelet_5pc}
\end{figure}

\begin{figure}
  \centerline{\includegraphics[width=14cm]{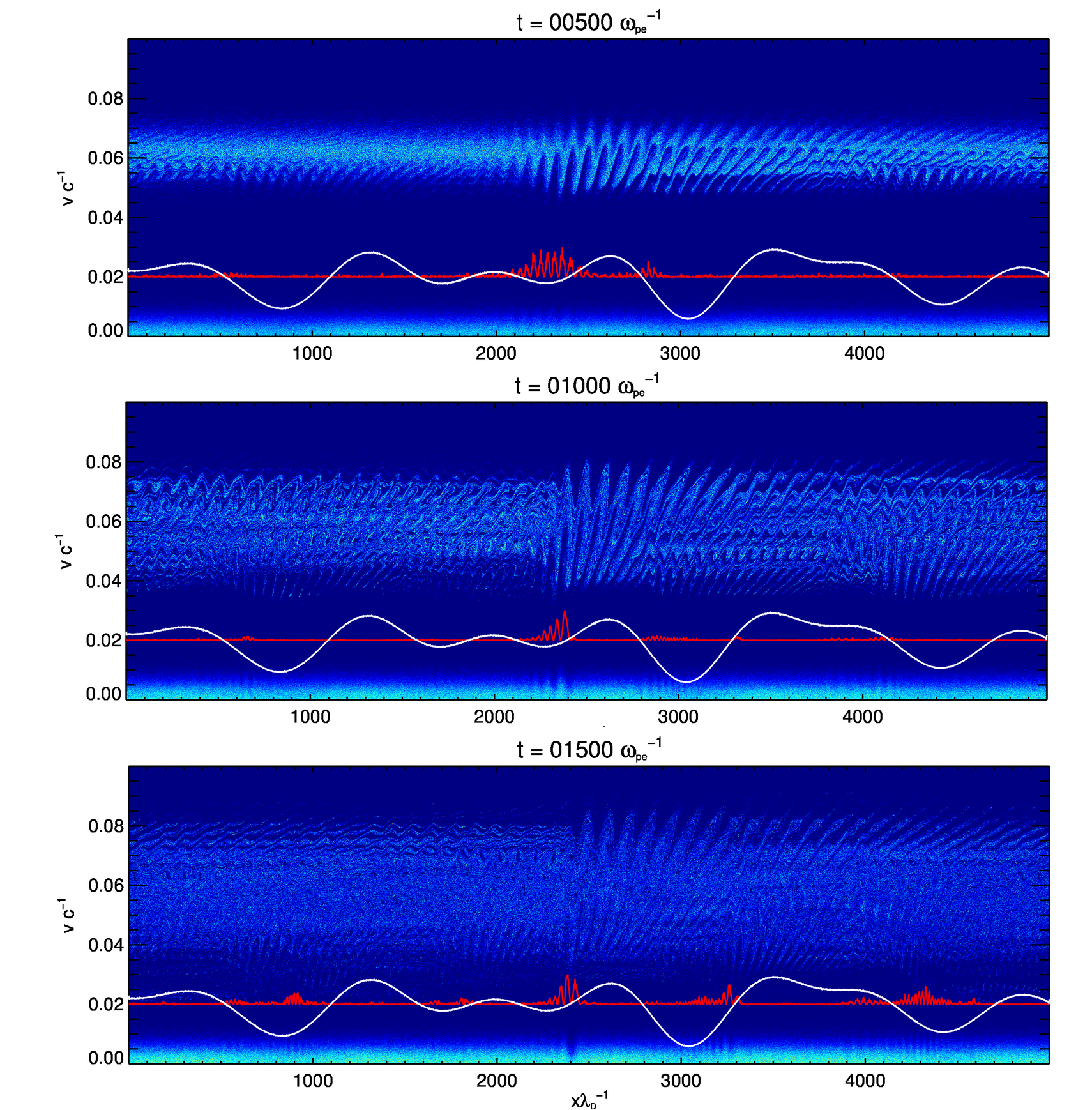}}
  \caption{Phase-space distribution background and beam electrons. The white curve illustrates the density profile, and the red curve the localisation of electric field energy. The turning points in the density profile / concentrations of electric field energy separate distinct regions of phase space, with the production of high-velocity particle  downstream of the turning points. Due to the periodic nature of the boundary, and the crossing times of the particles, these jets are quickly recycled through the boundary to the upstream regions and make multiple passes through the acceleration sites.}
  \label{phasespace_5pc}
\end{figure}

We now turn our attention to the five simulations where the 
average density fluctuation is in excess of the 
quasi-homogeneous threshold equation \ref{density_threshold}, 
and consider how they differ from
 the homogeneous case considered in Section \ref{homo_results}. 
Returning to Figure \ref{fig_vdf}, we note that all of the 
inhomogeneous simulations differ from the homogeneous 
case in that (i) the deceleration of the bulk of the 
beams particles and associate plateau formation is enhanced, 
and (ii) beam's particles are also accelerated to higher 
velocities leading to the formation of a high energy tail. 
Thus, we immediately see that there are two competing 
effects of the changes to wave-particle resonances due to 
the inhomogeneity, namely; (i) \textit{an enhanced deceleration process} 
due to a broader range of available particle-to-wave resonances, 
and (ii) \textit{an acceleration process} due to the re-absorption 
of wave-field energy into accelerated beam particles. 
The particular rates of these two process and the competition 
between them are found to vary in the inhomogeneous regime 
depending on the particular choice of ${\Delta}n$. 
The difference is more clear in 
the evolution of energy budgets 
(such as overall wave and particle energies in Figure \ref{fig_energy}, 
and  the beam kinetic energy budgets in different velocity ranges in
Figure \ref{fig_kebands}) 
and in the evolution of the wave-field spectrum 
(Figures \ref{kw_1pc}--\ref{kw_5pc}, and 
Figures \ref{packet_1pc}--\ref{packet_5pc}).

First considering the overall particle and wave energy as per 
Figure \ref{fig_energy}, we see that in all inhomogeneous cases considered
the net effect of the resonance broadening has been to 
significantly curtail the maximum level of gained by the waves 
(panel a), and thus lost by the particles (panel b). In comparison 
to the homogeneous case, the peak total wave-field energy is in 
the range $(5-10) \times 10^{-10}\,\mathrm{J}$. This is around $2-4$ 
times less than the homogeneous case, which peaks at around 
$1.9 \times 10^{-9}\,\mathrm{J}$. 
The peak wave energies (or minimum particle energies) are, 
in all cases, reached at earlier times than in the 
homogeneous case and so the overall instability timescale 
has been reduced (which is consistent with the more rapid 
plateau formation visible in Figure \ref{fig_vdf}).  
The initial growth rates leading to the maxima/minima of 
wave/particle energy are found to vary depending on the choice 
of ${\Delta}n$. For ${\Delta}n = 0.01, 0.02$ and $0.03$ growth 
rates are actually initially slightly faster than the 
homogeneous rate, for ${\Delta}n = 0.04$ the growth rate is 
approximately the same as the homogeneous rate, and only for 
${\Delta}n = 0.05$  we see a comparatively slower growth rate. 
This pattern is also seen  in the overall value of the 
maxima/minimum wave/particle energies - within the class 
of inhomogeneous runs, ${\Delta}n = 0.01,0.02$ and $0.03$ 
achieve the highest amount of wave-particle energy exchange, 
whereas ${\Delta}n = 0.04$  and ${\Delta}n = 0.05$ achieve less. 
Following the peak/minimum wave/particle energy, the dominant 
energy exchange process is reversed and so the 
wave-fields have a net loss of energy to the particles, 
which appears to asymptotically tend to the same value 
for all choices of ${\Delta}n>0$. 
This is consistent with the end-state velocity distribution 
function in Figure \ref{fig_vdf}, whereby all inhomogeneous 
runs have approximately the same phase-space profile at 
$t\omega_{pe}=10,000$. 

Returning to Figure \ref{fig_kebands}, we now compare the 
results in terms of the evolution of the kinetic energy 
distribution  of the beam particles. 
The overall trend for the inhomogeneous runs is 
consistent with our description of Figures \ref{fig_vdf} 
and \ref{fig_energy} - compared to the ${\Delta}n=0$ run 
we see a enhanced deceleration of the thermal beam 
electrons (panel a, where $K_{p}\left(v<v_{b}+{\Delta}v_{b}\right)$).
 Not only is this a more rapid deceleration, but it is 
 continuous for sufficiently long to reach a greater 
 level of depletion, with less kinetic energy being 
 contained within the plateau at end times. However, 
 this is \textit{superficially} inconsistent with what we 
 have seen with Figure \ref{fig_energy}, whereby the 
 ${\Delta}n>0$ cases retain more overall particle energy 
 and transfer less overall energy to the Langmuir waves. 
 The resolution is, of course, in that the enhanced energy 
 loss due to deceleration is adequately compensated by 
 the acceleration associated with the high energy tail 
 formation, where we see a growth of kinetic energy 
 associated with particles in excess of the beam's thermal 
 range (panel b, where $K_{p}\left(v>v_{b}+{\Delta}v_{b}\right)$). 
 By late times, the combined net effect of the deceleration 
 and acceleration processes is such that the beam kinetic 
 energy for ${\Delta}n>0$ is in the range $0.87$--$0.88\,K_{p}(t=0)$, 
 far in excess of the ${\Delta}n=0$ case where it is 
 approximately $0.64\,K_{p}(t=0)$. Thus, in the inhomogeneous 
 case, we see that the beams loose less overall energy to 
 the background wave field. All choices of ${\Delta}n>0$ have 
 a similar end state in the thermal range (panel a), which is 
 consistent with the observation of Figure \ref{fig_vdf} that 
 all inhomogeneous runs considered appear to tend towards 
 the same plateau shape, height, and extent. For higher 
 choices of inhomogeneity ${\Delta}n=0.04$ and $0.05$, we 
 see in panel (b) that the acceleration process has 
 an earlier onset, specifically from the range (iii). 
 Interestingly, it appears that by the end state, the 
 acceleration to the largest velocities (panel c,  
 $K_{p}\left(v>v_{b}+3{\Delta}v_{b}\right)$)  is less efficient 
 for the ${\Delta}n=0.01$ and $0.05$ runs than the 
 ${\Delta}n=0.02,\,0.03,\,0.04$ runs, suggesting that this 
 process may not exhibit a simple dependence on the 
 amplitude of the fluctuations.

We now refer to Figures \ref{kw_1pc} and \ref{kw_5pc} which 
consider the evolution of the specific cases ${\Delta}n=0.01$ and ${\Delta}n=0.05$ 
respectively in $\left(k,\omega\right)$-space, and 
also Figure \ref{fig_kspe}, which shows the time-evolution 
across $k$-space (having integrated over $\omega$) for all values 
of ${\Delta}n$; on the same plot for ease of comparison.
All values of ${\Delta}n$ begin with an initial growth 
of spectral energy density concentrated in a peak 
about $k=0.072$, the expected value of prescribed by 
the resonance condition for the electron beam and the 
Langmuir wave dispersion relation for the case of a 
constant background density. 
Subsequently, from around $t{\omega}_{pe}=600$ onwards, 
we find that unlike in the ${\Delta}n=0$ case (where this 
peak grows to around $-170\,\mathrm{dB}$), for ${\Delta}n>0$ 
that the spectral energy density is smeared across a 
broader range, extending towards smaller values of $k$ due to 
the resonance broadening with a lower peak spectral 
energy density of around $-180\,\mathrm{dB}$.

As time further progresses, this broadening of the 
spectrum extends into negative-$k$ space.
In the homogeneous case the dispersion relation diagram 
shows that when resonant condition, 
$\sqrt{ \omega_{pe}^2 +3 V_{th,e}^2 k^2 } = {v_b}\,{k}$, 
of Langmuir waves and electron beam is met 
initially Langmuir wave packet is generated. 
Subsequently, due to quasi-linear relaxation 
$v_b$ essentially decreases as it joins the bulk distribution
therefore for the right-hand-side of the latter equality to stay the same,
 $k$ needs increase. 
This behaviour can be clearly seen in
Figures \ref{fig_kspe} and \ref{kw_0pc}. 
In the homogeneous case no electron acceleration is seen, as expected.
In the case with ${\Delta}n=0.01$  we see
appearance of negative wavenumbers $k$ along the expected dispersion curve (Figure \ref{kw_1pc}). This implies that 
backwards propagating Langmuir waves appear. Because parametric instabilities such as electrostatic decay (see e.g., \citealt{1985PhR...129..285Z}) are inhibited due to immobile ions, 
only one possibility is allowed --
the appearance of negative wavenumbers $k$ (Figures \ref{fig_kspe} and \ref{kw_1pc}) 
can be attributed to 
the Langmuir wave refraction on the positive density 
gradient parts of the inhomogeneity. 
As shown by \citet{2014PhPl...21a2903P} for the case of Langmuir waves propagating on a single large-scale density gradient, on positive density 
gradient regions  $k$ deceases,
which implies that Langmuir waves are resonant with higher  velocity electrons, 
$V_{ph}= \omega_{Langmuir} / k$,
hence we see particle acceleration in the inhomogeneous case, but no 
acceleration in the uniform density case. 
Because the dispersion relation 
parabola is shallow, even a small increase in number 
density can lead to a large decrease in $k$.
The drift towards smaller $k$, including negative $k$, does not 
require three-wave interaction or ion-sound-mediated electrostatic decay and 
naturally occurs due to wave refraction alone.
The negative density gradient and quasi-linear relaxation 
both lead to increase in $k$, 
but do not result in the particle acceleration.
For large density fluctuations with ${\Delta}n=0.05$, 
 the dispersion relation diagram  (see Figure \ref{kw_5pc}) shows
all of the features of the ${\Delta}n=0.01$ case, however Langmuir wave 
packets  slide off the Langmuir dispersion 
parabola because of non-linear modifications in a similar manner as 
described by \citet{2015A&A...584A..83T}. Note that for the time-window used the resulting frequency resolution of the discrete Fourier transform is  ${\Delta}\omega = 0.0672\,{\omega}_{pe}$, which is of the order of the separation of the power concentration from the linear dispersion relationship, with them at most being separated by a few points. As such, the significance of this separation is unclear in these figures. We find that if larger time-windows are chosen (of say thousands of $\omega_{pe}^{-1}$), the separation becomes clearly resolved, confirming that the feature is not just an artefact. However, such figures have too low a time-resolution to show the procession to smaller-$k$ (the key feature) and as such are not shown here.
Comparing amongst different inhomogeneous simulations, 
it is clear that the progression to negative-$k$ (i.e., 
the amount of Langmuir wave reflection) scales with ${\Delta}n$, 
whereby the largest amplitude inhomogeneous profile  
${\Delta}n=0.05$ (\textit{viz. the profile with the steepest gradients}) 
produces a population of backwards-propagating waves most rapidly, 
out-pacing less steep profiles, contrasting with the most gentle 
profile ${\Delta}n=0.01$ which reflects waves over a significantly 
longer time-scale.
This kinematic effect relating to the behaviour 
of Langmuir waves propagating in a longitudinally inhomogeneous 
medium can be observed directly in our simulations; 
specifically, the reflections in the density cavities are observed in the evolution of the electric field waveforms 
in $\left(x,t\right)$-space (Figures \ref{packet_1pc}, \ref{packet_5pc}, and \ref{td_5pc}), the co-spatial mode drift to low-$k$ modes is observed in wavelet power spectra of the electric field (Figure \ref{wavelet_5pc}) and the concomitant acceleration of particles is seen in phase space (Figure \ref{phasespace_5pc})  - all of which will be further discussed shortly.

We also find in Figures \ref{kw_1pc}, \ref{kw_5pc} and \ref{fig_kspe} that the ${\Delta}n>0$ spectra undergoes more rapid 
broadening towards higher positive-$k$ than in the ${\Delta}n=0$ 
case, and that this proceeds more rapidly for higher values of 
${\Delta}n$. In Section \ref{homo_results}, we attributed the 
shift to higher $k$ to a dynamic shift in the beam-particle 
resonance condition as the beam looses particles to the plateau 
formation. The observed rise in broadening rate as ${\Delta}n$ 
increases is thus a consequence of the enhanced deceleration 
of the bulk of the beam particles (i.e., more rapid plateau formation) 
for ${\Delta}n>0$ as we have previously commented on in our 
discussion of  Figures \ref{fig_vdf}, \ref{fig_energy} and \ref{fig_kebands}. 
Although the rate of the progression towards higher 
positive-$k$ is dependent on ${\Delta}n$, we find that 
at approaching the end time the extent of the 
spectrum towards positive-$k$ is approximately the same 
for all runs, which is consistent with the similar end 
states of the ${\Delta}n>0$ beam's in phase space 
(see Figure \ref{fig_vdf}).

Next, we consider the Langmuir wave behaviour in $(x,t)$-space, 
shown in Figures \ref{packet_1pc} and \ref{packet_5pc}, 
which for the specific case of ${\Delta}n=0.01$ and ${\Delta}n=0.05$
show  the evolution of the electric 
field in $(x,t)$-space in the case (blue curve), a 
scaled-measure of electric field energy density ($10E_{X}^{2}-0.5$, 
red-curve), and the particular density profile (black curve) (i.e., they are equivalent to Figure \ref{packet_0pc} for the homogeneous case). 
In all cases we observe an obvious localisation effect in which 
the wave fields and their energy densities become concentrated 
in the local density cavities, whereby the highest peaks often 
coincide with locations of maximum local gradient (corresponding 
to Langmuir wave turning points). This high degree of 
localisation leads to a clumping of energy density 
many orders of magnitude above background levels. We also 
note that this magnitude is not necessarily greater for 
greater ${\Delta}n$, although the larger choices of ${\Delta}n$ 
result in more local density cavities with high gradients, 
and so we see a greater number of localised peaks, or clumps, 
for higher ${\Delta}n$. This behaviour is completely different 
to the  ${\Delta}n=0$ case shown in Figure \ref{packet_0pc} 
(discussed in Section \ref{homo_results}). 
In the movies, one can see that the energy 
density localisations are propagating, although they
typically remain trapped in the cavity (but they 
do \textit{not} sit at a singular point, such as the 
local density minima or the local gradient maxima). 
In some cases, as these clumps propagate to the up 
the walls of a cavity, clear reflections at the local 
gradient maxima (which act as a turning point) can 
be observed (in particular, see the case of ${\Delta}n=0.05$ 
whereby many of the individual density cavities 
posses a steep gradient). Thus, we can directly observe the formation of a 
population of counter-propagating Langmuir waves 
due to Langmuir wave reflection off inhomogeneities 
which has been previously-discussed in the context 
of the spectral results.  Additionally, one can directly 
observe the loss of wave energy density to the 
high-energy particle acceleration process. As an example, 
consider the $ {\Delta}n=0.01$ movie near the point 
$x=1200{\lambda}_{D}$ from times $t\omega_{pe}=2000-4000$. 
As the concentration propagates up along the cavity, it 
becomes trapped at the maxima and looses energy density rapidly. 
These reflections can also be clearly demonstrated without appealing to animation by considering time-distance diagrams of the electric field energy, as shown in Figure \ref{td_5pc} for the case ${\Delta}n=0.05$. In Figure \ref{td_5pc}, the reflections of waves within density cavities, associated intensification on turning points, and crossings of counter-propagating wave fronts can all be readily observed.  Note, the levels chosen to produce the colour table are logarithmically spaced to enhance contrast between features (i.e., the figure provides information about the timings and location of reflections, and it is necessary to refer to Figure \ref{packet_5pc} for quantitative information regarding the field amplitudes.) 

The aforementioned decrease in $k$ seen in the Fourier spectra, which we earlier state can only be due to this refraction process, can be further confirmed by considering the evolution of wavelet power spectra of the electric fields, which allows the concentration of wave power in $\left(k,x\right)$-space to be examined, i.e., unlike Fourier transforms, we can examine the distribution $k$ in $x$.  
Figure \ref{wavelet_5pc} shows the concentration of wavelet power at instances in time for the ${\Delta}n=0.05$ case (specifically, the figure shows the root of wavelet power, for best contrast). The white curve illustrates the density profile in $x$, and the straight white line is the homogeneous beam-particle resonance $k{\lambda_{D}}=0.072$. The black, labelled contour indicates the $95\%$ confidence level in the signal (see \citet{Torrence98apractical} for details on wavelets and their statistical properties).
Crucially, the figure demonstrates that the conversion to low-$k$ modes occurs in the locality of the turning points - confirming our earlier description of the process (compare  Figure \ref{td_5pc} with the reflections in Figure \ref{td_5pc}). This qualitative behaviour is common to all inhomogeneous cases considered, and is absent in the ${\Delta}n=0$ case. The figure also demonstrates that some wave packets are (temporarily) converted to higher-$k$ modes during their propagation down negative density gradients. The movie clearly shows that the low-$k$ power emanates from the turning points on the density gradients, and also that the production of low-$k$ power occurs in a bursty fashion, which suggests that corresponding particle acceleration may also occur in bursts. 

Finally, the particle acceleration concomitant to the refraction can be seen in the electron phase space, shown for the case of ${\Delta}n=0.05$ in Figure \ref{phasespace_5pc}. Clear qualitative differences in the beam phase space can be seen upstream and downstream of the turning points in the density gradients can be seen, with the production of high-velocity particle jets in the downstream regions. As the propagation speed of free-streaming particles is rapid compared to $L_{x}$, the jets and other features cross through multiple density cavities and are rapidly recycled through the periodic boundary from the downstream region to the right to the upstream region to the left, making pinpointing particular features in phase-space difficult for our specific setup (although we stress that the periodic recycling is appropriate for our problem, which is intended to model the global properties of a beam passing through many density cavities). Further, due to disk-space restrictions and to reduce read/write operations, we only recorded the full velocity distribution (not integrated over $x$)  every $100\,\omega_{pe}^{-1}$, and as such cannot directly confirm whether the bursty behaviour of the production of low-$k$ waves does indeed lead to a bursty particle acceleration. 

These direct observations of the main physical aspects of the particle acceleration process, 
underpinning the changes of the beam-plasma instability in 
the inhomogeneous regime, are elaborated upon further in 
Section \ref{discussion}.

\section{Discussion}\label{discussion}

We have demonstrated in our PIC simulations that the 
introduction of an inhomogeneous density profile 
significantly affects the evolution of the electron 
beam-plasma instability.  The variable density profile 
with amplitudes in excess of the threshold equation \ref{density_threshold} apparently facilitates two processes that will compete to 
determine the overall beam energy loss and relaxation time, 
namely; (i) \textit{an enhanced deceleration process} and thus 
quicker plateau formation, and (ii) \textit{an acceleration process} 
which permits the formation of a high energy tail.

The enhanced deceleration process occurs due to a 
broader range of available particle-to-wave resonances.   
The acceleration process is associated with the re-absorption of 
wave-field energy into accelerated beam particles. This process is directly 
facilitated due to the refraction of Langmuir waves as they 
propagate along a region with a positive density gradient.  
There is clear evidence in Section \ref{inhomo_results} that 
as Langmuir waves propagate through regions of positive 
density gradient they experience a decrease in their 
wave-number $k$ (see, e.g., the wavelet power spectrum, Figure \ref{wavelet_5pc}). Thus, their phase velocity $v_{ph} = \omega/k$ increases 
and so the waves exchange energy resonantly with increasingly 
energetic particles during their propagation up the density gradients, producing high-velocity particle jets which can be seen in phase-space (Figure \ref{phasespace_5pc}).
The same process has been detailed for a beam propagating on a 
single large-scale positive density gradient by \citet{2014PhPl...21a2903P}.

The net effect of the two above processes was found in all 
inhomogeneous cases considered to be as follows; in 
comparison to the homogeneous case the bump in tail 
instability time-scale is actually quicker, with a faster 
procession towards plateau formation, although the beam 
population looses much less energy overall due to the 
high-energy tail formation.
We find no evidence of the 
slight positive slopes remaining in the distribution (integrated over the domain)
after saturation as discussed by \citet{2013AnGeo..31.1379V}, 
rather all cases considered here appear to relax to a plateau.
We also examined the local distribution functions at the end-time (i.e., at specific cells in the domain) to look for evidence of localised positive slopes, however it becomes clear that the number of PPC leads to a somewhat noisy distribution on the slope which could well obscure slightly positive slopes. We believe it is likely that the formation of the gentle positive slopes is a subtle kinetic effect which is shown in the semi-analytical models of  \citet{2013AnGeo..31.1379V}, but would require very high PPC to demonstrate with a PIC simulation.

We also find that all cases considered in excess of the inhomogeneity amplitude threshold tend to relax to a similar asymptotic end-state, both in terms of the distribution function (Fig \ref{fig_vdf}) and energy distribution of particles (Fig \ref{fig_kebands}). This result is consistent with the work of \citet{2015ApJ...807...38V}, who considered inhomogeneous beam-plasma interaction with a probabilistic model (see e.g., their Figure 4). Our results are consistent as both studies do not venture outside a quasi-linear / weak beam regime, both are 1D and both ignore the influence of ion-sound waves (in our case, due to ion immobilisation). At this stage, we cannot make any comments on an underlying explanation for this remarkable result, beyond noting that we have obtained the same with a PIC approach, and highlighting that the particular end state was shown to be dependent on the initial beam energy and beam density in the \citet{2015ApJ...807...38V} study.
The end state, consisting of  an asymptotic equilibrium reached between the Langmuir waves and the suprathermal electrons, has been explored in the case of kappa-distributed plasmas by \citet{2011PhPl...18l2303Y,2012PhPl...19a2304Y,2012PhPl...19e2301Y}. It is possible that future PIC simulations similar to those considered in this paper could provide insight into the nature of the asymptotic equilibrium, by varying parameters other than the density profile to alter the end-state,  and also give an indication as to whether Gaussian- or Kappa-distributions are more plausible equilibrium distribution functions.

In addition to the overall effect of the density 
cavities on the beam's energy loss and instability 
saturation time it is also interesting that some 
individual density cavities assume steep enough 
gradients such that Langmuir packets are 
directly observed to reflect about turning points (e.g., Figure \ref{td_5pc}). 
This shows that simple Langmuir wave kinematics in the 
inhomogeneous regime facilitates the development of a 
population of counter propagating Langmuir waves. 
The presence of a population of counter propagating 
Langmuir waves is an essential component in typical 
models of harmonic solar radio emission whereby their 
three-wave coalescence produces the emission. 
Usually, the population of counter-propagating Langmuir 
waves is thought to be created through electrostatic 
decay of the forward propagating waves into 
ion-acoustic waves and backwards propagating 
waves (such as in the self-consistent 
emission simulations of, e.g., 
\citealt{2015A&A...584A..83T}). However, 
in this case we find that a population 
of counter-propagating Langmuir waves can 
easily be created in the absence of such 
three-wave interactions, a population 
which could then proceed to coalesce and 
participate in harmonic radio emission. It should be noted that beating Langmuir waves in the absence of ion-acoustic counterparts have actually been occasionally observed in-situ  (e.g., \citealt{1999JGR...10417069K}).
It would be interesting to compare radio emission 
resulting from the homogeneous and inhomogeneous 
cases in future work, although it is likely using a 
full PIC approach will suffer severe computational 
demands (for example, the relatively expensive 1D simulations 
considered here would need to be extended into two 
dimensions to study radio emission). 

As stressed in Section \ref{setup}, these experiments have specifically considered the influence of \lq{strong}\rq{} inhomogeneity, i.e., that in excess of the fluctuation-amplitude threshold Equation (\ref{density_threshold}). We have not considered the intermediate case of inhomogeneity levels less than, or approaching, this value due to computational restrictions. Given that our PIC results broadly confirm the results of \citet{2013ApJ...778..111K} in the strongly inhomogeneous case, we find no reason to suggest that the physics and end-state of the intermediate cases should be different to those found using their model, in which cases of weak inhomogeneity not satisfying the threshold simply proceed as per the perfectly homogeneous case.

\section{Conclusions}\label{conclusion}

We have for the first time shown with a fully kinetic PIC approach 
that plasma inhomogeneity can seriously alter behaviour of the 
electron beam-plasma instability, independently confirming previous 
theoretical predictions and modelling efforts with a clear, 
robust and simple numerical approach. 

We find that in the case of no fluctuations (homogeneous density), the dispersion relation diagram 
shows that when the resonant condition, $\sqrt{ \omega_{pe}^2 +3 V_{th,e}^2  k^2 } = {v_b}\,{k}$, 
of Langmuir waves and electron beam is met, initially Langmuir wave packets are generated. 
Then, because of quasi-linear relaxation (when beam particles join the main 
population forming the plateau) $v_b$ essentially decreases as it joins the bulk distribution
therefore for the right-hand-side of the latter equality to stay the same,
(i.e. for wave packet to 
stay on the dispersion curve) $k$ needs to increase. This behaviour can be clearly seen in
Figures \ref{fig_kspe} and \ref{kw_0pc}. In the homogeneous case no electron acceleration is seen, as expected.
In the case of ${\Delta}n=0.01$, the dispersion 
relation (Figure \ref{kw_1pc}) 
shows appearance of negative wavenumbers. This implies that 
backwards propagating Langmuir waves appear. 
Because parametric instabilities
are inhibited due to immobile ions, 
there is only one possibility:
the appearance of negative wavenumbers (Figures \ref{kw_1pc} and \ref{fig_kspe}) can be attributed to 
the Langmuir wave refraction on the positive density gradient parts of the inhomogeneity.  We have directly observed this refraction in the electric field waveforms and corresponding wavelet power spectra.
As discussed by \citet{2014PhPl...21a2903P}, at positive density gradient locations $k$ deceases,
which means that Langmuir waves are resonant with higher  velocity electrons, 
$V_{ph}= \omega_{Langmuir} / k$,
hence we see particle acceleration in the inhomogeneous case, but no 
acceleration in the uniform density case. 
Because the dispersion relation 
parabola is shallow, even small increase in number 
density can lead to large decrease in $k$ (even making it negative).
The key point  is that drift towards smaller $k$, including negative does not 
require three-wave interaction or ion-sound-mediated decay and 
naturally occurs due to wave refraction alone.
(1) negative density gradient and (2) quasi-linear relaxation both lead to increase in $k$, 
but do not result in the particle acceleration.
For the higher-level fluctuations considered,
 e.g. ${\Delta}n=0.05$, 
 the dispersion relation diagram  (see Figure \ref{kw_5pc}) shows
all the features of the ${\Delta}n=0.01$ case, however Langmuir wave 
packets now slide off the Langmuir dispersion 
parabola because of non-linearity, in a similar manner as 
described by \citet{2015A&A...584A..83T}.

In conclusion, fully kinetic PIC simulations broadly confirm findings 
of quasi-linear and the Hamiltonian model based on
Zakharov's equations with a kinetic treatment of the beam only. 
We also find that the strong density fluctuations 
(e.g, 5 \% of the background) modify properties
of excited Langmuir waves altering their dispersion properties.

\bibliography{references}

\begin{thebibliography}{32}
\providecommand{\natexlab}[1]{#1}
\providecommand{\url}[1]{\texttt{#1}}
\expandafter\ifx\csname urlstyle\endcsname\relax
  \providecommand{\doi}[1]{doi: #1}\else
  \providecommand{\doi}{doi: \begingroup \urlstyle{rm}\Url}\fi

\bibitem[{Ginzburg} and {Zhelezniakov}(1958)]{1958SvA.....2..653G}
V.~L. {Ginzburg} and V.~V. {Zhelezniakov}.
\newblock {On the Possible Mechanisms of Sporadic Solar Radio Emission
  (Radiation in an Isotropic Plasma)}.
\newblock \emph{\sovast}, 2:\penalty0 653, October 1958.

\bibitem[{Forslund} et~al.(1975){Forslund}, {Kindel}, {Lee}, {Lindman}, and
  {Morse}]{1975PhRvA..11..679F}
D.~W. {Forslund}, J.~M. {Kindel}, K.~{Lee}, E.~L. {Lindman}, and R.~L. {Morse}.
\newblock {Theory and simulation of resonant absorption in a hot plasma}.
\newblock \emph{\pra}, 11:\penalty0 679--683, February 1975.
\newblock \doi{10.1103/PhysRevA.11.679}.

\bibitem[{Kim} et~al.(2007){Kim}, {Cairns}, and
  {Robinson}]{2007PhRvL..99a5003K}
E.-H. {Kim}, I.~H. {Cairns}, and P.~A. {Robinson}.
\newblock {Extraordinary-Mode Radiation Produced by Linear-Mode Conversion of
  Langmuir Waves}.
\newblock \emph{Physical Review Letters}, 99\penalty0 (1):\penalty0 015003,
  July 2007.
\newblock \doi{10.1103/PhysRevLett.99.015003}.

\bibitem[{Reid} and {Ratcliffe}(2014)]{2014RAA....14..773S}
H.~A.~S. {Reid} and H.~{Ratcliffe}.
\newblock {A review of solar type III radio bursts}.
\newblock \emph{Research in Astronomy and Astrophysics}, 14:\penalty0 773-804,
  July 2014.
\newblock \doi{10.1088/1674-4527/14/7/003}.

\bibitem[{Sturrock}(1964)]{1964NASSP..50..357S}
P.~A. {Sturrock}.
\newblock {Type III Solar Radio Bursts}.
\newblock \emph{NASA Special Publication}, 50:\penalty0 357, 1964.

\bibitem[{Lin} et~al.(1981){Lin}, {Potter}, {Gurnett}, and
  {Scarf}]{1981ApJ...251..364L}
R.~P. {Lin}, D.~W. {Potter}, D.~A. {Gurnett}, and F.~L. {Scarf}.
\newblock {Energetic electrons and plasma waves associated with a solar type
  III radio burst}.
\newblock \emph{\apj}, 251:\penalty0 364--373, December 1981.
\newblock \doi{10.1086/159471}.

\bibitem[{Anderson} et~al.(1981){Anderson}, {Eastman}, {Gurnett}, {Frank}, and
  {Parks}]{1981JGR....86.4493A}
R.~R. {Anderson}, T.~E. {Eastman}, D.~A. {Gurnett}, L.~A. {Frank}, and G.~K.
  {Parks}.
\newblock {Plasma waves associated with energetic particles streaming into the
  solar wind from the earth's bow shock}.
\newblock \emph{\jgr}, 86:\penalty0 4493--4510, June 1981.
\newblock \doi{10.1029/JA086iA06p04493}.

\bibitem[{Gurnett} et~al.(1978){Gurnett}, {Anderson}, {Scarf}, and
  {Kurth}]{1978JGR....83.4147G}
D.~A. {Gurnett}, R.~R. {Anderson}, F.~L. {Scarf}, and W.~S. {Kurth}.
\newblock {The heliocentric radial variation of plasma oscillations associated
  with type III radio bursts}.
\newblock \emph{\jgr}, 83:\penalty0 4147--4152, September 1978.
\newblock \doi{10.1029/JA083iA09p04147}.

\bibitem[{Muschietti} et~al.(1995){Muschietti}, {Roth}, and
  {Ergun}]{1995JGR...10017481M}
L.~{Muschietti}, I.~{Roth}, and R.~E. {Ergun}.
\newblock {Kinetic localization of beam-driven Langmuir waves}.
\newblock \emph{\jgr}, 100:\penalty0 17481--17490, September 1995.
\newblock \doi{10.1029/95JA00595}.

\bibitem[{Muschietti} et~al.(1996){Muschietti}, {Roth}, and
  {Ergun}]{1996JGR...10115605M}
L.~{Muschietti}, I.~{Roth}, and R.~E. {Ergun}.
\newblock {On the formation of wave packets in planetary foreshocks}.
\newblock \emph{\jgr}, 101:\penalty0 15605--15614, July 1996.
\newblock \doi{10.1029/96JA00926}.

\bibitem[{Ryutov}(1969)]{1969JETP...30..131R}
D.~D. {Ryutov}.
\newblock {Quasilinear Relaxation of an Electron Beam in an Inhomogeneous
  Plasma}.
\newblock \emph{Soviet Journal of Experimental and Theoretical Physics},
  30:\penalty0 131, 1969.

\bibitem[{Breizman} and {Ryutov}(1970)]{1970JETPL..11..421B}
B.~N. {Breizman} and D.~D. {Ryutov}.
\newblock {Influence of Inhomogeneity of Plasma on the Relaxation of an
  Ultrarelativistic Electron Beam}.
\newblock \emph{Soviet Journal of Experimental and Theoretical Physics
  Letters}, 11:\penalty0 421, June 1970.

\bibitem[Nishikawa and Ryutov(1976)]{doi:10.1143/JPSJ.41.1757}
Kyoji Nishikawa and D.~D. Ryutov.
\newblock Relaxation of relativistic electron beam in a plasma with random
  density inhomogeneities.
\newblock \emph{Journal of the Physical Society of Japan}, 41\penalty0
  (5):\penalty0 1757--1765, 1976.
\newblock \doi{10.1143/JPSJ.41.1757}.
\newblock URL \url{http://dx.doi.org/10.1143/JPSJ.41.1757}.

\bibitem[{Voshchepynets} and {Krasnoselskikh}(2013)]{2013AnGeo..31.1379V}
A.~{Voshchepynets} and V.~{Krasnoselskikh}.
\newblock {Electron beam relaxation in inhomogeneous plasmas}.
\newblock \emph{Annales Geophysicae}, 31:\penalty0 1379--1385, August 2013.
\newblock \doi{10.5194/angeo-31-1379-2013}.

\bibitem[{Krafft} et~al.(2013){Krafft}, {Volokitin}, and
  {Krasnoselskikh}]{2013ApJ...778..111K}
C.~{Krafft}, A.~S. {Volokitin}, and V.~V. {Krasnoselskikh}.
\newblock {Interaction of Energetic Particles with Waves in Strongly
  Inhomogeneous Solar Wind Plasmas}.
\newblock \emph{\apj}, 778:\penalty0 111, December 2013.
\newblock \doi{10.1088/0004-637X/778/2/111}.

\bibitem[{Krafft} et~al.(2014){Krafft}, {Volokitin}, {Krasnoselskikh}, and {de
  Wit}]{2014JGRA..119.9369K}
C.~{Krafft}, A.~S. {Volokitin}, V.~V. {Krasnoselskikh}, and T.~D. {de Wit}.
\newblock {Waveforms of Langmuir turbulence in inhomogeneous solar wind
  plasmas}.
\newblock \emph{Journal of Geophysical Research (Space Physics)}, 119:\penalty0
  9369--9382, December 2014.
\newblock \doi{10.1002/2014JA020329}.

\bibitem[{Krafft} et~al.(2015){Krafft}, {Volokitin}, and
  {Krasnoselskikh}]{2015ApJ...809..176K}
C.~{Krafft}, A.~S. {Volokitin}, and V.~V. {Krasnoselskikh}.
\newblock {Langmuir Wave Decay in Inhomogeneous Solar Wind Plasmas: Simulation
  Results}.
\newblock \emph{\apj}, 809:\penalty0 176, August 2015.
\newblock \doi{10.1088/0004-637X/809/2/176}.

\bibitem[{Voshchepynets} et~al.(2015){Voshchepynets}, {Krasnoselskikh},
  {Artemyev}, and {Volokitin}]{2015ApJ...807...38V}
A.~{Voshchepynets}, V.~{Krasnoselskikh}, A.~{Artemyev}, and A.~{Volokitin}.
\newblock {Probabilistic Model of Beam-Plasma Interaction in Randomly
  Inhomogeneous Plasma}.
\newblock \emph{\apj}, 807:\penalty0 38, July 2015.
\newblock \doi{10.1088/0004-637X/807/1/38}.

\bibitem[{Voshchepynets} and {Krasnoselskikh}(2015)]{2015JGRA..12010139V}
A.~{Voshchepynets} and V.~{Krasnoselskikh}.
\newblock {Probabilistic model of beam-plasma interaction in randomly
  inhomogeneous solar wind}.
\newblock \emph{Journal of Geophysical Research (Space Physics)}, 120:\penalty0
  10, December 2015.
\newblock \doi{10.1002/2015JA021705}.

\bibitem[Arber et~al.(2015)Arber, Bennett, Brady, Lawrence-Douglas, Ramsay,
  Sircombe, Gillies, Evans, Schmitz, Bell, and Ridgers]{Arber:2015hc}
T~D Arber, K~Bennett, C~S Brady, A~Lawrence-Douglas, M~G Ramsay, N~J Sircombe,
  P~Gillies, R~G Evans, H~Schmitz, A~R Bell, and C~P Ridgers.
\newblock {Contemporary particle-in-cell approach to laser-plasma modelling}.
\newblock \emph{Plasma Physics and Controlled Fusion}, 57\penalty0
  (11):\penalty0 1--26, November 2015.

\bibitem[Melrose and McPhedran(1991)]{9780511600036}
D.~B. Melrose and R.~C. McPhedran.
\newblock \emph{Electromagnetic Processes in Dispersive Media}.
\newblock Cambridge University Press, 1991.
\newblock ISBN 9780511600036.
\newblock URL \url{http://dx.doi.org/10.1017/CBO9780511600036}.
\newblock Cambridge Books Online.

\bibitem[{Thurgood} and {Tsiklauri}(2015)]{2015A&A...584A..83T}
J.~O. {Thurgood} and D.~{Tsiklauri}.
\newblock {Self-consistent particle-in-cell simulations of fundamental and
  harmonic plasma radio emission mechanisms}.
\newblock \emph{\aap}, 584:\penalty0 A83, December 2015.
\newblock \doi{10.1051/0004-6361/201527079}.

\bibitem[{Ratcliffe} et~al.(2014){Ratcliffe}, {Brady}, {Che Rozenan}, and
  {Nakariakov}]{2014PhPl...21l2104R}
H.~{Ratcliffe}, C.~S. {Brady}, M.~B. {Che Rozenan}, and V.~M. {Nakariakov}.
\newblock {A comparison of weak-turbulence and particle-in-cell simulations of
  weak electron-beam plasma interaction}.
\newblock \emph{Physics of Plasmas}, 21\penalty0 (12):\penalty0 122104,
  December 2014.
\newblock \doi{10.1063/1.4904065}.

\bibitem[{Lotov} et~al.(2015){Lotov}, {Timofeev}, {Mesyats}, {Snytnikov}, and
  {Vshivkov}]{2015PhPl...22b4502L}
K.~V. {Lotov}, I.~V. {Timofeev}, E.~A. {Mesyats}, A.~V. {Snytnikov}, and V.~A.
  {Vshivkov}.
\newblock {Note on quantitatively correct simulations of the kinetic
  beam-plasma instability}.
\newblock \emph{Physics of Plasmas}, 22\penalty0 (2):\penalty0 024502, February
  2015.
\newblock \doi{10.1063/1.4907223}.

\bibitem[{Baumg{\"a}rtel}(2013)]{2013AnGeo..31..633B}
K.~{Baumg{\"a}rtel}.
\newblock {Coherent amplitude modulation of electron-beam-driven Langmuir
  waves}.
\newblock \emph{Annales Geophysicae}, 31:\penalty0 633--638, April 2013.
\newblock \doi{10.5194/angeo-31-633-2013}.

\bibitem[{Zakharov} et~al.(1985){Zakharov}, {Musher}, and
  {Rubenchik}]{1985PhR...129..285Z}
V.~E. {Zakharov}, S.~L. {Musher}, and A.~M. {Rubenchik}.
\newblock {Hamiltonian approach to the description of non-linear plasma
  phenomena}.
\newblock \emph{\physrep}, 129:\penalty0 285--366, December 1985.
\newblock \doi{10.1016/0370-1573(85)90040-7}.

\bibitem[{Pechhacker} and {Tsiklauri}(2014)]{2014PhPl...21a2903P}
R.~{Pechhacker} and D.~{Tsiklauri}.
\newblock {Three-dimensional particle-in-cell simulation of electron
  acceleration by Langmuir waves in an inhomogeneous plasma}.
\newblock \emph{Physics of Plasmas}, 21\penalty0 (1):\penalty0 012903, January
  2014.
\newblock \doi{10.1063/1.4863494}.

\bibitem[Torrence and Compo(1998)]{Torrence98apractical}
Christopher Torrence and Gilbert~P. Compo.
\newblock A practical guide to wavelet analysis.
\newblock \emph{Bulletin of the American Meteorological Society}, 79:\penalty0
  61--78, 1998.

\bibitem[{Yoon}(2011)]{2011PhPl...18l2303Y}
P.~H. {Yoon}.
\newblock {Asymptotic equilibrium between Langmuir turbulence and suprathermal
  electrons}.
\newblock \emph{Physics of Plasmas}, 18\penalty0 (12):\penalty0 122303,
  December 2011.
\newblock \doi{10.1063/1.3662105}.

\bibitem[{Yoon}(2012{\natexlab{a}})]{2012PhPl...19a2304Y}
P.~H. {Yoon}.
\newblock {Asymptotic equilibrium between Langmuir turbulence and suprathermal
  electrons in three dimensions}.
\newblock \emph{Physics of Plasmas}, 19\penalty0 (1):\penalty0 012304, January
  2012{\natexlab{a}}.
\newblock \doi{10.1063/1.3676159}.

\bibitem[{Yoon}(2012{\natexlab{b}})]{2012PhPl...19e2301Y}
P.~H. {Yoon}.
\newblock {Electron kappa distribution and steady-state Langmuir turbulence}.
\newblock \emph{Physics of Plasmas}, 19\penalty0 (5):\penalty0 052301, May
  2012{\natexlab{b}}.
\newblock \doi{10.1063/1.4710515}.

\bibitem[{Kellogg} et~al.(1999){Kellogg}, {Goetz}, {Monson}, and
  {Bale}]{1999JGR...10417069K}
P.~J. {Kellogg}, K.~{Goetz}, S.~J. {Monson}, and S.~D. {Bale}.
\newblock {Langmuir waves in a fluctuating solar wind}.
\newblock \emph{\jgr}, 104:\penalty0 17069--17078, August 1999.
\newblock \doi{10.1029/1999JA900163}.

\end{thebibliography}

\subsection*{Acknowledgements}
The authors acknowledge funding from the Leverhulme Trust Research 
Project Grant RPG-311. The computational work for this paper was carried 
out on the joint BIS, STFC and SFC (SRIF) funded DiRAC-1 cluster at the 
University of St. Andrews (Scotland, UK). 
The EPOCH code used in this work was in part funded by the UK EPSRC grants 
EP/G054950/1, EP/G056803/1, EP/G055165/1 and EP/ M022463/1. Wavelet software was provided by C. Torrence and G. Compo, and is available at URL: http://paos.colorado.edu/research/wavelets/  The authors would also like to thank the anonymous referees for their time spent reviewing the manuscript, and their constructive criticisms and input.

\end{document}